\documentclass[reqno,12pt]{amsart}

\usepackage{fullpage}
\usepackage{graphicx}
\usepackage{amssymb}
\usepackage{amsmath}
\usepackage{cite}
\usepackage{epstopdf}

\numberwithin{equation}{section}
\def\ie/{i.e.}
\def\eg/{e.g.}
\def\etc/{etc.}
\def\cf/{cf.}

\def\u{\tilde{u}}
\def\V{\nu}
\def\g{\tilde{g}}
\def\f{\tilde{f}}

\def\L{\tilde{L}}

\def\c{{\rm c}}

\def\E{{\mathcal E}}
\def\P{{\mathcal P}}

\def\C{\tilde{C}}

\def\arctanh{\,{\rm arctanh}}

\def\tbinom#1#2{{\textstyle\binom{#1}{#2}}}
\def\const{\text{const.}}

\newtheorem{prop}{Proposition}

\def\scrpt#1{$\scriptstyle {#1}$}

\begin{document}
\allowdisplaybreaks[3]

\title{Travelling waves and conservation laws for\\ 
highly nonlinear wave equations\\
modelling Hertz chains}

\author{
Michelle Przedborski$^1$ 
\lowercase{\scshape{and}}
Stephen C. Anco$^2$
\\\\\lowercase{\scshape{
${}^1$Department of Physics\\
Brock University\\
St. Catharines, ON L\scrpt2S\scrpt3A\scrpt1, Canada}} 
\\\\
\lowercase{\scshape{
${}^2$Department of Mathematics and Statistics\\
Brock University\\
St. Catharines, ON L\scrpt2S\scrpt3A\scrpt1, Canada}}
}

\begin{abstract}
A highly nonlinear, fourth-order wave equation that models 
the continuum theory of long wavelength pulses 
in weakly compressed, discrete, homogeneous chains 
with a general power-law contact interaction 
is studied. 
For this wave equation, 
all solitary wave solutions and all nonlinear periodic wave solutions, 
along with all conservation laws, 
are derived. 
The solutions are explicitly parameterized in terms of 
the asymptotic value of the wave amplitude in the case of solitary waves
and the peak of the wave amplitude in the case of nonlinear periodic waves. 
All cases in which the solution expressions can be stated in an explicit analytic form 
using elementary functions are worked out. 
In these cases, 
explicit expressions for the total energy and total momentum for all solutions 
are obtained as well. 
The derivation of the solutions 
uses the conservation laws combined with an energy analysis argument 
to reduce the wave equation directly to a separable first-order differential equation 
which determines the wave amplitude in terms of the travelling wave variable. 
This method can be applied more generally to other highly nonlinear wave equations. 
\end{abstract}

\maketitle

\begin{center}\small
email: 
mp06lj@brocku.ca, 
sanco@brocku.ca
\end{center}

\section{Introduction}\label{sec:intro}

Travelling waves are an interesting feature of many nonlinear phenomena
and have numerous important physical applications. 
One area that has attracted much recent attention is the study of 
compression pulses in one-dimensional systems of discrete macroscopic particles 
that interact by a power-law contact potential 
\cite{Nesterenko1983,Nesterenko1985,Nesterenko1994,Sinkovits1995,Nesterenko1995,Sen1996, Coste1997, Sen1998, Chatterjee1999, Hinch1999, Hong1999, Ji1999,Manciu1999a,Manciu1999b, Sen1999, Hascoet2000, Manciu2001, Nesterenko2001,Sen2001, Rosas2003, Rosas2004, Daraio2005, English2005, Nesterenko2005, Daraio2006,Job2007, Sokolow2007, Zhen2007, Porter2008, Sen2008, Herbold2009, Porter2009, Rosas2010,Santibanez2011,James2012, Khatri2012, Stefanov2012, Takato2012, Vitelli2012}.
When such a discrete system is pre-compressed, 
then propagating solitary waves and nonlinear periodic waves 
can be produced through suitable tuning of the initial conditions
\cite{Nesterenko2001,Daraio2006, Sen2008, Herbold2009, James2012}. 
Pulses whose wavelength is much larger than the particle radius, 
but much smaller than the length of the chain, 
can be modeled by a continuum partial differential equation (PDE). 
Continuum models are important because 
they allow comparison with experimental findings 
and give insights into general features of compression waves 
in the underlying discrete system. 

The simplest model describes a homogeneous chain of 
$N>2$ spherical particles with mass $m$
such that the interaction contact potential is given by the Hertz law~\cite{Hertz1882} 
$V=a\delta^{5/2}$  
where $a$ is a constant 
which depends on the material properties of the particles,
and $\delta=\delta_0+u_{i}-u_{i+1}\geq 0$ is the (dynamical) overlap distance 
between adjacent particles, 
given in terms of the relative particle positions $u_{j}$, $j=1,\ldots,N$, 
and the initial overlap $\delta_0$. 
This model has an interesting generalization in which 
the exponent in the Hertz potential is replaced by an arbitrary power, 
$V=a\delta^{k+1}$, 
where $k>1$ for typical physical applications
\cite{Spence1968,Johnson1985,Johnson2005}. 
When the initial overlap $\delta_0$ is small compared to the particle displacements, 
the chain is said to be weakly compressed. 
In this case, 
the continuum PDE for the relative displacement $u = -\u(x,t)$ of the particles 
is given by the highly nonlinear, fourth-order wave equation
\cite{Nesterenko1994,Porter2008,Porter2009,Nesterenko2001}
\begin{equation}\label{eq:tilu-pde}
c^{-2} \u_{tt} = \u_x^{k-1}\u_{xx} + \alpha \u_x^{k-3}\u_{xx}^3 + \beta \u_x^{k-2}\u_{xx}\u_{xxx} + \gamma \u_x^{k-1}\u_{xxxx}
\end{equation}
with
\begin{gather}
c = \sqrt{k(k+1) a(2R)^{k+1}/m}, 
%\sqrt{k(k+1)}\frac{\sqrt{2a}(R_1+R_2)^{\frac{k+1}{2}}}{\sqrt{m_1 + m_2}}
\label{eq:tilu-pde-c}
\\
\alpha =\tfrac{1}{6}(k-1)(k-2) R^2,
%\frac{m_1^2}{(m_1 + m_2)^2}
%G
\quad
\beta =\tfrac{2}{3}(k-1) R^2,
%\frac{2m_1-m_2}{m_1 + m_2}
%H
\quad
\gamma = \tfrac{1}{3} R^2,
%\frac{m_1^2 - m_1 m_2 + m_2^2}{(m_1+m_2)^2}
%I 
\label{eq:tilu-pde-coeffs}
\end{gather}
where 
$R$ is the radius of the particles. 
Note, in these expressions \eqref{eq:tilu-pde-c}--\eqref{eq:tilu-pde-coeffs}, 
the constants $\alpha$, $\beta$, $\gamma$ have units of length-squared,
and the constant $c$ has units of speed,
while $\u$ has units of length. 

An important remark is that the unitless strain 
\begin{equation}\label{eq:v-strain}
v = \u_x =-u_x > 0
\end{equation}
rather than the displacement $\u=-u$, 
will be the variable whose dynamics will physically describe a propagating wave.
The strain is positive because the underlying discrete dynamical system has no restoring forces. 
For solitary waves, the strain will have the asymptotic behaviour 
\begin{equation}\label{eq:v-solitarywave-bc}
v\to v_0 >0 \quad\text{ as }\quad x\to \pm\infty
\end{equation}
where $v_0$ is a constant background strain. 
The boundary conditions on the strain for nonlinear periodic waves 
consist of
\begin{equation}\label{eq:v-periodicwave-bc}
v(x,t) = v(x+L,t+T),
\quad
v_x(x,t) = v_x(x+L,t+T)
\end{equation}
where $L$ is the wavelength and $T$ is the period. 

Existence of solitary wave solutions for the highly nonlinear wave equation \eqref{eq:tilu-pde} 
has been proved recently \cite{Stefanov2012}. 
However, 
an exact analytical expression for these solutions has not appeared 
in the literature to-date. 
Moreover, 
the only known exact solution is a special periodic wave \cite{Porter2009} 
that contains no free parameters. 

In the present paper, 
we derive exact analytical expressions for 
all solitary wave solutions and all nonlinear periodic solutions of 
this wave equation \eqref{eq:tilu-pde}. 
We also obtain analytic expressions for their total energy and total momentum. 
The solutions are parameterized in terms of 
the asymptotic value of the wave amplitude in the case of solitary waves
and the peak of the wave amplitude in the case of nonlinear periodic waves. 

Our method involves the following steps. 
First, we find the conservation laws admitted by the wave equation \eqref{eq:tilu-pde}. 
Next, we use these conservation laws to derive a first-order ordinary differential equation (ODE) for all travelling waves, without having to do any explicit integrations. 
This reduction contains two free constants which correspond to first integrals 
related to the energy flux and longitudinal-momentum flux of a travelling wave. 
Third, we employ an energy analysis argument to classify the types of travelling waves. 
This argument shows that there are three types of solutions:
(i) solitary waves with an exponentially decaying tail, 
which satisfy the asymptotic condition \eqref{eq:v-solitarywave-bc};
(ii) periodic waves, which satisfy the periodicity condition \eqref{eq:v-periodicwave-bc};
(iii) cusped waves, both periodic and solitary;
(iv) nodal waves, in which the wave amplitude vanishes at isolated points. 
Fourth, we solve the ODE in each of these different solution cases 
by giving an explicit integral expression that determines the wave amplitude 
in terms of the travelling wave variable. 
Last, we show how to use the ODE along with its first integrals 
to obtain the total energy and total momentum carried by these solutions, 
without the need to evaluate complicated integrals involving the explicit form of the solutions. 

The paper is organized as follows. 

In section~\ref{sec:prelims}, 
we first present the Lagrangian and Hamiltonian formulations of 
the highly nonlinear wave equation \eqref{eq:tilu-pde}, 
and we next derive its conservation laws and corresponding conserved quantities
by a direct method which does not require the use of Noether's theorem. 
This material also contains some general results which are applicable to 
the study of any fourth-order wave equation.

In section~\ref{sec:soln-steps} 
we apply the steps in our method to the wave equation \eqref{eq:tilu-pde}
for an arbitrary exponent $k>1$. 
This yields explicit integral expressions for 
solitary waves, nonlinear periodic waves, cusped solitary waves, cusped periodic waves, 
and nodal waves, 
which are the main results of the paper. 
We also derive conserved integral expressions for the total energy and total momentum 
carried by these different kinds of travelling waves. 

In section~\ref{sec:examples}, 
we systematically determine all cases of the power-law exponent $k>1$ 
for which each of the different solution integrals 
can be evaluated in an explicit analytical form using elementary functions. 
This analysis yields 
two explicit solitary waves, for $k=2,3$; 
two explicit periodic waves, 
one for $k=2$ with zero energy-flux and non-negative longitudinal momentum-flux, 
and the other for $k=3$ with non-negative energy-flux and zero longitudinal momentum-flux; 
three explicit cusped waves, 
two being cusped solitary waves for $k=2,3$,
and the third being a cusped periodic wave 
for $k=3$ with zero longitudinal momentum-flux; 
two explicit nodal waves, 
one for $k=2$ with non-negative longitudinal momentum-flux, 
and the other for $k>1$ arbitrary with zero longitudinal momentum-flux. 
We give explicit expressions for the total energy and total momentum 
for each of these solutions. 

Finally, we make some concluding remarks in section~\ref{sec:remarks}.

\section{Preliminaries}\label{sec:prelims}

For the highly nonlinear wave equation \eqref{eq:tilu-pde}, 
we will first present its Lagrangian formulation 
which we will then use in interpreting its conservation laws.

\subsection{Variational formulation}

We start by considering a general fourth-order wave equation 
$u_{tt} = F(u,u_x,u_{xx},u_{xxx},$ $u_{xxxx})$. 
An equation of this form has a Lagrangian formulation 
if it arises from the extremals of a Lagrangian function 
$L=\frac{1}{2}u_t^2 +L_0(u,u_x,u_{xx},u_{xxx},u_{xxxx})$ with respect to $u$. 
A Lagrangian function is not unique, 
because adding any total derivative expression to it 
does not change its extremals. 
This freedom can be used to find a Lagrangian of lowest possible order. 

The extremals of $L$ are obtained by setting 
its variational derivative to be zero,
where the variational derivative (Euler-Lagrange operator) is defined by~\cite{Olver2000,Bluman2010}
\begin{equation}
\frac{\delta}{\delta u} = 
\frac{\partial}{\partial u} 
-D_t \frac{\partial}{\partial u_t} 
-D_x \frac{\partial}{\partial u_x} 
+ D_x^2 \frac{\partial}{\partial u_{xx}} 
+ D_t^2 \frac{\partial}{\partial u_{tt}} 
+ D_tD_x \frac{\partial}{\partial u_{tx}} 
+\cdots 
\end{equation}
which has the property that it annihilates any total derivative expression. 
Here $D$ denotes a total derivative (acting by the chain rule), namely
\begin{equation}
\begin{aligned}
& D_t = 
\frac{\partial}{\partial t}+ u_t\frac{\partial}{\partial u} 
+ u_{tt} \frac{\partial}{\partial u_t} 
+u_{tx} \frac{\partial}{\partial u_x} 
+\cdots,
\\
& D_x = 
\frac{\partial}{\partial x}+ u_x\frac{\partial}{\partial u} 
+ u_{tx} \frac{\partial}{\partial u_t} 
+u_{xx} \frac{\partial}{\partial u_x} 
+\cdots .
\end{aligned}
\end{equation}
Hence, 
a Lagrangian formulation then consists of finding a function $L_0$ for which 
\begin{equation}\label{eq:general-L}
0=u_{tt} - F = -\frac{\delta L}{\delta u} = u_{tt} -\frac{\delta L_0}{\delta u}.
\end{equation}
It is straightforward to show that, since $F$ is fourth-order, 
if $L_0$ contains any terms of higher than second order 
then these terms must take the form of a total derivative, 
as otherwise they will produce terms of higher than fourth order 
in the variational derivative $\delta L_0/\delta u$. 
Consequently, 
there is no loss of generality if we require $L_0$ to be only second order.

Thus, a Lagrangian formulation \eqref{eq:general-L} will exist 
for $u_{tt} = F(u,u_x,u_{xx},u_{xxx},u_{xxxx})$ 
if (and only if) we can find a function $L_0(u,u_x,u_{xx})$ that satisfies
\begin{equation}
\frac{\delta L_0}{\delta u} = 
\frac{\partial L_0}{\partial u} 
-D_x \frac{\partial L_0}{\partial u_x} 
+ D_x^2 \frac{\partial L_0}{\partial u_{xx}} 
= F.
\end{equation}

The highly nonlinear wave equation \eqref{eq:tilu-pde}
has a simple Lagrangian function 
\begin{equation}\label{eq:L} 
L=\tfrac{1}{2}\u_t^2 +L_0(\u_x,\u_{xx}),
\quad
L_0 = c^2\big(  \tfrac{-1}{k(k+1)} \u_x^{k+1} + \tfrac{1}{6}R^2 \u_x^{k-1} \u_{xx}^2 \big). 
\end{equation}
This Lagrangian has the lowest possible order. 
By comparison, in Ref.~\cite{Nesterenko2001} 
a more complicated higher-order Lagrangian was derived. 
It is easy to check that the difference between that Lagrangian 
and the one given here is just a total $x$-derivative expression.

Next we show how to convert this Lagrangian formulation into 
an equivalent Hamiltonian formulation using the strain variable \eqref{eq:v-strain},
as this will be helpful later for physically interpreting 
the conservation laws admitted by the highly nonlinear wave equation \eqref{eq:tilu-pde}.

We consider a general Lagrangian wave equation of the form \eqref{eq:general-L}.
This equation can be written equivalently as 
a first-order system in time derivatives 
if we introduce variables for the first-order derivatives of $u$:
\begin{equation}
u_t=w,
\quad
u_x=-v.
\end{equation}
Then we have $w_t= u_{tt} = F = \delta L_0/\delta u$, 
and $v_t= -u_{tx}= -w_x$. 
The variational derivative $\delta L_0/\delta u$ 
on the righthand side of the $w_t$ equation 
can be expressed in terms of $v$ through the identity~\cite{Olver2000}
\begin{equation}
\frac{\delta }{\delta u} = D_x \frac{\delta }{\delta v}.
\end{equation}
Using this identity, we now have $w_t= D_x(\delta L_0/\delta v)$. 
The righthand side of the $v_t$ equation can be expressed in a similar form
by noting $w_x = D_x(\delta(\tfrac{1}{2}w^2)/\delta w)$,
so then we have $v_t= -D_x(\delta(\tfrac{1}{2}w^2)/\delta w)$.
Thus, the original wave equation becomes 
\begin{equation}\label{eq:general-vw-sys}
\begin{pmatrix} v_t \\ w_t \end{pmatrix}
= \mathcal{D} \begin{pmatrix} \delta \mathcal{H}/\delta v \\ \delta \mathcal{H}/\delta w \end{pmatrix},
\end{equation}
where 
\begin{equation}\label{eq:general-H}
\mathcal{H} = \int_{-\infty}^{\infty} H\,dx,
\quad
H = \tfrac{1}{2}w^2 -L_0
\end{equation}
is the Hamiltonian,
and where
\begin{equation}\label{eq:general-D}
\mathcal{D} = \begin{pmatrix} 0 & -D_x \\ -D_x & 0 \end{pmatrix}
\end{equation}
is a skew-adjoint matrix operator. 
(Note the choice of sign in $\mathcal{D}$ corresponds to a positive sign 
for the kinetic term $\tfrac{1}{2}w^2=\tfrac{1}{2}u_t^2$ in $H$). 
There is a non-canonical Poisson bracket associated to this formulation. 
For any two functionals 
$\mathcal{F} = \int_{-\infty}^{\infty} F\,dx$ and $\mathcal{G} = \int_{-\infty}^{\infty} G\,dx$,
given in terms of functions $F(v,w,v_x,w_x,v_{xx},w_{xx},\ldots)$ and $G(v,w,v_x,w_x,v_{xx},w_{xx},\ldots)$, 
the Poisson bracket is defined by
\begin{equation}
\{\mathcal{F},\mathcal{G}\} = 
\int_{-\infty}^{\infty} \begin{pmatrix} \delta \mathcal{F}/\delta v \\ \delta \mathcal{F}/\delta w \end{pmatrix}^{\mathrm{t}} \mathcal{D} \begin{pmatrix} \delta \mathcal{G}/\delta v \\ \delta \mathcal{G}/\delta w \end{pmatrix}\,dx 
\end{equation}
where ``$\mathrm{t}$'' denotes the matrix transpose.
This bracket obeys the standard antisymmetry and Jacobi properties~\cite{Olver2000},
due to the skew-adjoint form of $\mathcal{D}$ 
combined with the property that the components of $\mathcal{D}$ 
have no dependence on the variables $v$ and $w$. 
One immediate consequence of a Hamiltonian formulation is that 
the Hamiltonian is a constant of motion, namely
\begin{equation}
\frac{d\mathcal{H}}{dt} = 0 
\end{equation}
whenever $v$ and $w$ obey suitable asymptotic decay conditions for large $x$. 

The Hamiltonian for the highly nonlinear wave equation \eqref{eq:tilu-pde}
is given by 
\begin{equation}\label{eq:weak-Hamil} 
H = \tfrac{1}{2}w^2 + c^2\big(  \tfrac{-1}{k(k+1)} v^{k+1} + \tfrac{1}{6}R^2 v^{k-1} v_{x}^2 \big). 
\end{equation}
This Hamiltonian will physically describe a conserved energy, 
consisting of a kinetic energy term $\tfrac{1}{2}w^2$ 
plus a potential energy term. 

It may also be of interest to note that this Hamiltonian formulation 
can be re-expressed as a second-order equation in time derivatives for the strain $v$. 
Consider the general Hamiltonian system \eqref{eq:general-vw-sys}--\eqref{eq:general-D}. 
By taking the time derivative of the $v$ equation and substituting the $w$ equation, 
we obtain $v_{tt}= -w_{tx} = D_x^2(\delta\mathcal{H}/\delta v)$.
This yields a wave equation with the form 
\begin{equation}
v_{tt}= (\delta H_0/\delta v)_{xx}, 
\quad
H_0=-L_0,
\end{equation}
which is neither Lagrangian nor Hamiltonian in terms of $v$. 
Here $H_0$ can be interpreted as the potential energy. 
For the highly nonlinear wave equation \eqref{eq:tilu-pde}, 
we have 
\begin{equation}
H_0 = c^2\big(  \tfrac{-1}{k(k+1)} v^{k+1} + \tfrac{1}{6}R^2 v^{k-1} v_{x}^2 \big)
\end{equation}
which gives 
\begin{equation}\label{eq:weak-v-eqn}
v_{tt}= c^2\big(  \tfrac{-1}{k(k+1)} v^{k+1} + \tfrac{1}{6}R^2 v^{k-1} v_{x}^2 \big){}_{xx}.
\end{equation}

We will next derive the conservation laws and corresponding conserved quantities
for the highly nonlinear wave equation \eqref{eq:tilu-pde}.

\subsection{Conservation laws}\label{sec:ConsLaws}

The most commonly used method for finding conservation laws of 
Lagrangian differential equations is Noether's theorem \cite{Olver2000,Bluman2010}. 
This method produces a conservation law from every symmetry of a Lagrangian. 
For example, 
if a Lagrangian is invariant to within a total divergence under time translations, 
then this symmetry typically produces the conservation law for energy. 
Some drawbacks to using Noether's theorem are that, first, 
the symmetries of the Lagrangian need to be determined,
and second, the variation of the Lagrangian under each symmetry needs to be 
found to obtain the expression for the resulting conservation law. 
We will instead use an alternative, modern version~\cite{Bluman2010,Anco-review} of Noether's theorem,
employing multipliers rather than symmetries, 
which avoids the usual drawbacks and which is simpler to carry out. 

To explain the method and to review the theory of conservation laws, 
we consider a general fourth-order wave equation 
$u_{tt} =F(u,u_x,u_{xx},u_{xxx},u_{xxxx})$. 
A conservation law for a wave equation of this form 
consists of a local continuity equation 
\begin{equation}\label{eq:conslaw}
D_t T + D_x X =0
\end{equation}
holding for all solutions $u(t,x)$ of the wave equation, 
where $T$ is the conserved density and $X$ is the spatial flux,
which are given by functions of $t$, $x$, $u$, $u_t$, and $x$-derivatives of $u$ and $u_t$. 
Note that $u_{tt}$ (and any derivatives of it) can be assumed to be eliminated
from $T$ and $X$ through the wave equation. 

Every conservation law \eqref{eq:conslaw} yields a conserved quantity 
defined by the global continuity equation
\begin{equation}
\frac d{dt} \int_{\Omega} T\, dx = -X\Big|_{\partial\Omega}
\end{equation}
on a given spatial domain $\Omega$, 
with endpoints $\partial\Omega$. 
For solutions $u(t,x)$ of the wave equation having suitable boundary conditions 
at the endpoints of the domain, 
the spatial flux terms will vanish so then the quantity 
\begin{equation}\label{eq:C}
C[u]= \int_{\Omega} T\, dx=\const
\end{equation}
is a constant of motion. 
Two conservation laws are physically equivalent 
if they give the same conserved quantity up to boundary terms. 
This occurs if (and only if) their conserved densities $T$ differ by 
a total $x$-derivative,  $D_x\Psi$, 
and correspondingly, their fluxes $X$ differ by 
a total time-derivative, $-D_t\Psi$, 
for all solutions $u(t,x)$ of the wave equation. 
Consequently, a conservation law will be physically trivial if, 
on all solutions, the density and flux have the form 
$T= D_x\Psi$ and $X= - D_t\Psi$
whereby the local continuity equation \eqref{eq:conslaw} will hold trivially.
Thus, equivalent conservation laws differ by a trivial conservation law. 

Any conservation law \eqref{eq:conslaw} has an equivalent characteristic form 
that holds as an identity obtained by moving off solutions of the wave equation:
\begin{equation}\label{eq:charform}
D_tT +D_x(X-\Phi)= (u_{tt} -F)Q,
\end{equation}
where 
\begin{equation}\label{eq:trivX}
\Phi = (u_{tt} -F)E_{u_{tx}}(T)+ D_x(u_{tt} -F)E_{u_{txx}}(T) + \cdots
\end{equation}
is a trivial flux which vanishes on all solutions of the wave equation, 
and where 
\begin{equation}\label{eq:TQrelation}
Q=E_{u_t}(T) 
\end{equation}
is called the multiplier. 
Here 
\begin{equation}
E_Z  =\frac{\partial}{\partial Z}-D_x\frac{\partial}{\partial Z_x} +D_x^2\frac{\partial}{\partial Z_{xx}} +\cdots
\end{equation}
denotes the (spatial) Euler-Lagrange operator with respect to a variable $Z$.
This operator has the property that $E_Z(f)$ vanishes precisely 
if $f$ is a total $x$-derivative of an expression in terms of $x$, $Z$, and $x$-derivatives of $Z$. 
We provide a short derivation of equations \eqref{eq:charform}--\eqref{eq:TQrelation} 
at the end of this section. 

Note that the characteristic equation \eqref{eq:charform} reduces to 
the conservation law \eqref{eq:conslaw} when $u$ satisfies the wave equation $u_{tt}=F$. 
This formulation of conservation laws is important for several reasons. 
First, it shows that there do not exist any non-trivial conserved densities $T$
that depend only on $t$, $x$, $u$, and $x$-derivatives of $u$, $u_t$. 
In particular, 
$T$ must have some essential dependence on $u_t$ (or its $x$-derivatives), 
as otherwise $D_t T +D_x X$ cannot vanish whenever $u$ satisfies $u_{tt}=F$.
Second, from this conclusion, 
it follows that if $Q=E_{u_{t}}(T)=0$ then $T=\tilde T+D_x\Psi$ holds 
where $\tilde T$ has no dependence on $u_t$ (and its $x$-derivatives), 
thereby implying $\tilde T=0$. 
This shows that the multiplier of a conservation law will be trivial 
precisely when the conserved density is trivial. 
Thus, there is a one-to-one relationship between non-trivial conserved densities (up to equivalence) and non-vanishing multipliers. 
Moreover, when $Q\neq 0$, the conserved density $T$ and the flux $X$ 
can be recovered from $Q$ by a direct integration of the characteristic equation \eqref{eq:charform}. 

This reduces the problem of finding conservation laws to the simpler problem of finding multipliers. 
From the characteristic equation \eqref{eq:charform}, 
we see that a function $Q$ will be a multiplier if (and only if) 
its product with the wave equation has the form of a total $t$-derivative plus a total $x$-derivative.
Such expressions have the characterization that their variational derivative
with respect to $u$ vanishes identically. 
This condition 
\begin{equation}\label{eq:Qeqn}
\frac{\delta}{\delta u}\Big( (u_{tt} -F)Q \Big)=0
\end{equation}
will split into a system of linear partial differential equations that determine $Q$. 
First, from relation \eqref{eq:TQrelation}, 
all multipliers are functions only of $t$, $x$, $u$, $u_t$, and $x$-derivatives of $u$ and $u_t$. 
Then, since $u_{tt}$, $u_{ttx}$, \etc/ do not appear in $Q$, 
the coefficients of these variables in the condition \eqref{eq:Qeqn} 
must vanish separately. 
The resulting equations give an overdetermined system that can be 
straightforwardly solved for $Q$. 
We emphasize that this system is applicable whether or not a Lagrangian formulation exists for the wave equation $u_{tt}=F$. 

In the case when a wave equation $u_{tt}=F(u,u_x,u_{xx},u_{xxx},u_{xxxx})$ 
possesses a Lagrangian formulation \eqref{eq:general-L}, 
the determining system for multipliers $Q$ can be simplified 
using some tools from variational calculus. 
This leads to the system 
\begin{gather}
\hat D_t^2 Q =F_u Q + F_{u_x}D_x Q + F_{u_{xx}}D_x^2 Q + F_{u_{xxx}}D_x^3 Q + F_{u_{xxxx}}D_x^4 Q, 
\label{eq:symm}\\
Q_{u_t} = E_{u_t}(Q),
\quad
Q_{u_{tx}} = -E_{u_t}^{(1)}(Q),
\quad
Q_{u_{txx}} = E_{u_t}^{(2)}(Q),
\quad
\ldots,
\label{eq:Helmholtz}
\end{gather}
where 
\begin{equation}\label{eq:Dt-solns}
\hat D_t = D_t\big|_{u_{tt}=F} 
= \frac{\partial}{\partial t}+ u_t\frac{\partial}{\partial u} 
+ F \frac{\partial}{\partial u_t} 
+ u_{tx} \frac{\partial}{\partial u_x} 
+D_x F \frac{\partial}{\partial u_{tx}}
+ u_{txx} \frac{\partial}{\partial u_{xx}} 
+\cdots
\end{equation}
is the total time derivative evaluated on the space of solutions of the wave equation, 
and where
\begin{equation}
\begin{aligned}
& E_Z^{(1)} = \frac{\partial}{\partial Z_x}- \tbinom{2}{1}D_x\frac{\partial}{\partial Z_{xx}} +\tbinom{3}{1}D_x^2\frac{\partial}{\partial Z_{xxx}} +\cdots,
\\
& E_Z^{(2)} = \frac{\partial}{\partial Z_{xx}}-\tbinom{3}{2}D_x\frac{\partial}{\partial Z_{xxx}} +\tbinom{4}{2}D_x^2\frac{\partial}{\partial Z_{xxxx}} +\cdots,
\\
& \qquad \vdots
\end{aligned}
\end{equation}
are higher-order Euler-Lagrange operators~\cite{Olver2000,Anco-review}.
The first equation in the system \eqref{eq:symm}--\eqref{eq:Helmholtz} 
coincides with the determining equation 
for infinitesimal symmetries $\mathbf X=Q\partial/\partial_u$ of the wave equation,
with the symmetry defined to act on solutions $u(t,x)$ of the wave equation 
by the infinitesimal transformation 
$u \rightarrow u + \epsilon Q + O(\epsilon^2)$ 
in terms of a parameter $\epsilon$. 
The remaining equations in the system \eqref{eq:symm}--\eqref{eq:Helmholtz} 
comprise the Helmholtz conditions~\cite{Olver2000} which are necessary and sufficient for 
$Q$ to have the form of an Euler-Lagrange expression \eqref{eq:TQrelation}.
The Helmholtz conditions turn out to be equivalent to the condition that 
the Lagrangian of the wave equation is invariant to within a total derivative
under the prolonged symmetry 
${\rm pr}\mathbf X=Q\partial/\partial_u + D_x Q\partial/\partial_{u_x} + D_x^2 Q\partial/\partial_{u_{xx}} + \cdots$. 
Consequently, 
multipliers correspond directly to symmetries of the Lagrangian. 
This provides a simple formulation of Noether's theorem that avoids any use of the Lagrangian itself~\cite{Anco1997,Bluman2010,Anco-review}. 

For the highly nonlinear wave equation \eqref{eq:tilu-pde}, 
we will now present all conservation laws given by conserved densities of the first-order form 
\begin{equation}\label{eq:1storderT}
T(t,x,\u,\u_x,\u_{xx},\u_t)
\end{equation}
which can be expected to encompass all of the physically relevant conserved quantities 
\cite{Anco-review}. 
The corresponding multipliers will have the form
\begin{equation}\label{eq:1storderQ}
Q(t,x,\u,\u_x,\u_t)
\end{equation}
as shown by the variational relation \eqref{eq:TQrelation}. 
The results are obtained by directly solving the determining system \eqref{eq:symm}--\eqref{eq:Helmholtz}. 
For this class of multipliers \eqref{eq:1storderQ}, 
note that the characteristic equation simplifies to 
\begin{equation}\label{eq:1stordercharform}
D_tT +D_xX= (\u_{tt} -F)Q
\end{equation}
since the trivial flux term \eqref{eq:trivX} vanishes. 
We use this equation to obtain the conserved density and spatial flux 
corresponding to each multiplier. 
Specifically, as $\u_{tt}$, $\u_{ttx}$, \etc/ do not appear in $T$, $X$, and $Q$, 
the coefficients of these variables in equation \eqref{eq:1stordercharform} 
must vanish separately, 
whereby this equation splits into a system of linear partial differential equations that are easily integrated to find explicit expressions for $T$ and $X$. 

All calculations for our results have been carried out using Maple.

The first-order multipliers admitted by the highly nonlinear wave equation \eqref{eq:tilu-pde}
are given by 
\begin{equation}\label{eq:Qs} 
Q_1 = 1, 
\quad
Q_2= t,
\quad
Q_3 = -\u_x,
\quad
Q_4 = \u_t . 
\end{equation}
The corresponding conserved density and flux determined by each multiplier 
are given by, respectively, 
\begin{align}
&\begin{aligned}
& T_1 = \u_t,
\\
& X_1 = -c^2\big( \tfrac{1}{3}R^2 \u_x^{k-1} \u_{xxx} + \tfrac{k-1}{6}R^2 \u_x^{k-2} \u_{xx}^2 + \tfrac{1}{k} \u_x^k \big); 
\end{aligned}
\label{eq:TX1}
\\
&\begin{aligned}
& T_2 = t \u_t - \u,
\\
& X_2 = -c^2 t \big( \tfrac{1}{3}R^2 \u_x^{k-1} \u_{xxx} + \tfrac{k-1}{6}R^2 \u_x^{k-2} \u_{xx}^2 + \tfrac{1}{k} \u_x^k \big); 
\end{aligned}
\label{eq:TX2}\\
& \begin{aligned}
& T_3= -\u_t \u_x,
\\
& X_3 = \tfrac{1}{2} \u_t^2 +c^2  \big( \tfrac{1}{3}R^2 \u_x^{k} \u_{xxx} + \tfrac{k-2}{6} R^2 \u_x^{k-1} \u_{xx}^2 + \tfrac{1}{k+1} \u_x^{k+1} \big); 
\end{aligned}
\label{eq:TX3}\\
& \begin{aligned}
& T_4 = \tfrac{1}{2} \u_t^2 - c^2 \big( \tfrac{1}{6}R^2 \u_x^{k-1} \u_{xx}^2 - \tfrac{1}{k(k+1)} \u_x^{k+1} \big), 
\\
& X_4 = -c^2\big( \tfrac{1}{3}R^2 \u_x^{k-1}(\u_t \u_{xxx} -\u_{tx} \u_{xx})  + \tfrac{k-1}{6}R^2 \u_x^{k-2}\u_t\u_{xx}^2 + \tfrac{1}{k} \u_x^k\u_t \big) . 
\end{aligned}
\label{eq:TX4}
\end{align}
These conservation laws yield the following conserved quantities:
\begin{align}
& C_1[\u]=\int_{\Omega} \u_t \,dx, 
\label{eq:C1}\\
& C_2[\u]=\int_{\Omega} t\u_t - \u \,dx  
= tC_1[\u] -\int_{\Omega} \u \,dx,  
\label{eq:C2}\\
& C_3[\u]= \int_{\Omega} -\u_t \u_x\,dx,
\label{eq:C3}\\
& C_4[\u]=\int_{\Omega} \big( \tfrac{1}{2} \u_t^2 - c^2 ( \tfrac{1}{6}R^2 \u_x^{k-1} \u_{xx}^2 - \tfrac{1}{k(k+1)} \u_x^{k+1} ) \big) \,dx. 
\label{eq:C4}
\end{align}
The quantity \eqref{eq:C4} coincides with the Hamiltonian \eqref{eq:weak-Hamil} 
which describes the physical total energy of wave solutions $u(t,x)$,
while the quantity \eqref{eq:C3} describes the physical total momentum of wave solutions. 
These interpretations are reinforced by the observation that the respective multipliers 
correspond to an infinitesimal time-translation $\mathbf X= \u_t\partial/\partial_{\u}$
and space-translation $\mathbf X= \u_x\partial/\partial_{\u}$
from the Lagrangian viewpoint. 

The quantities \eqref{eq:C1} and \eqref{eq:C2} have a less familiar physical meaning. 
We first note $\int_{\Omega} \u \,dx$ can be viewed as the mean amplitude 
associated to a wave solution $\u(t,x)$. 
Then we see 
\begin{equation}
\frac{d^2}{dt^2}\int_{\Omega} \u \,dx = \frac{dC_1[\u]}{dt} =0
\end{equation}
holds for wave solutions having sufficient asymptotic decay for large $x$. 
Thus the mean amplitude $\int_{\Omega} \u \,dx$ 
obeys the equation of free particle motion. 
Moreover, at the initial time $t=0$, 
the mean amplitude is related to the quantities \eqref{eq:C1} and \eqref{eq:C2} by 
\begin{equation}
\int_{\Omega} \u \,dx \Big|_{t=0} = -C_2[\u],
\quad
\frac{d}{dt}\int_{\Omega} \u \,dx \Big|_{t=0} = C_1[\u].
\end{equation}
Hence these quantities are just the initial value of the mean amplitude 
and the initial value of the mean longitudinal-momentum.

\subsection{Derivation of multiplier equation}

Here we give a short derivation of the characteristic equation \eqref{eq:charform}
for conservation laws of a general fourth-order wave equation 
$u_{tt} =F(u,u_x,u_{xx},u_{xxx},u_{xxxx})$. 

Since any conservation law \eqref{eq:conslaw} 
holds for all solutions of $u_{tt} =F$, 
we can re-write it using the time derivative \eqref{eq:Dt-solns} 
evaluated on solutions:
\begin{equation}
\hat D_tT +D_xX =0.
\end{equation}
Then, as $u_{tt}$ does not appear in this equation, 
it holds with $u(t,x)$ replaced by an arbitrary function. 
We now substitute the relation
\begin{equation}
\hat D_t - D_t = 
(F-u_{tt}) \frac{\partial}{\partial u_t} 
+D_x(F-u_{tt}) \frac{\partial}{\partial u_{tx}} 
+D_x^2(F-u_{tt}) \frac{\partial}{\partial u_{txx}} 
+\cdots, 
\end{equation}
which yields an identity
\begin{equation}\label{eq:hadamard}
D_tT +D_xX = (u_{tt} -F)K_0 + D_x(u_{tt} -F)K_1 + D_x^2(u_{tt} -F)K_2 + \cdots
\end{equation}
with 
\begin{equation}\label{eq:K}
K_0 = \frac{\partial T}{\partial u_t},
\quad
K_1 = \frac{\partial T}{\partial u_{tx}},
\quad
K_2 = \frac{\partial T}{\partial u_{txx}},
\quad
\ldots\  .
\end{equation}
Next we use integration by parts on each term on the righthand side 
in the identity \eqref{eq:hadamard}:
\begin{equation}
\begin{aligned}
D_x(u_{tt} -F)K_1 & = - (u_{tt} -F)D_xK_1 + D_x\big( (u_{tt} -F)K_1 \big),
\\
D_x^2(u_{tt} -F)K_2 & = (u_{tt} -F)D_x^2K_2 + D_x\big( D_x(u_{tt} -F)K_2 - (u_{tt} -F)D_xK_2 \big),
\\
& \quad\vdots
\end{aligned}
\end{equation}
Collecting the total derivative terms, we get 
\begin{equation}\label{eq:QKrelation}
\begin{aligned}
D_tT +D_xX & = 
(u_{tt} -F)\big( K_0 - D_xK_1 +D_x^2K_2 +\cdots) 
\\&\qquad
+ D_x\big( (u_{tt} -F)(K_1 -D_xK_2 +\cdots) + D_x(u_{tt} -F)(K_2 +\cdots) \big)
+\cdots .
\end{aligned}
\end{equation}
After substituting expressions \eqref{eq:K} into this equation, 
we obtain the characteristic equation \eqref{eq:charform} 
together with the expressions \eqref{eq:TQrelation} and \eqref{eq:trivX} 
for the multiplier and the trivial flux term.

\section{Travelling wave solutions}\label{sec:soln-steps} 

For a general fourth-order wave equation 
$u_{tt} =F(u,u_x,u_{xx},u_{xxx},u_{xxxx})$, 
the form of a traveling wave is $u = f(\xi)$ with $\xi = x-\V t$, 
where the constant $\V$ is the velocity of the wave,
and the function $f$ is the wave profile. 
Since $u_x = f'$ and $u_t = -\V f' = -\V u_x$, 
where a prime denotes the derivative with respect to $\xi$,
the wave equation will reduce to a fourth-order ODE 
\begin{equation}\label{eq:4thorderODE}
\V^2 f'' = F(f,f',f'',f''',f'''') 
\end{equation}
for $f(\xi)$. 

We will now explain how to use conservation laws 
to directly provide first integrals of the traveling wave ODE
and also to obtain conserved quantities 
for the solitary wave solutions and nonlinear periodic solutions of this ODE. 

First, 
suppose a conservation law $D_t T+D_x X=0$ of the wave equation $u_{tt}=F$ 
does not contain the variables $t$ and $x$ explicitly. 
Then this conservation law will give rise to a first integral of the traveling wave ODE \eqref{eq:4thorderODE}
by the reductions 
\begin{equation}\label{eq:reduced-derivatives}
D_t\big|_{u=f(\xi)} = -\V \frac{d}{d\xi},
\quad
D_x\big|_{u=f(\xi)} = \frac{d}{d\xi},
\end{equation}
yielding
\begin{equation}\label{eq:reduced-conslaw}
\frac{d}{dt} \Big( (X - \V T)\big|_{u=f(\xi)} \Big) = 0. 
\end{equation}
The first integral is thus given by 
\begin{equation}\label{eq:firstintegral}
X- \V T = C = \const . 
\end{equation}
It has the physical meaning of the spatial flux 
in a reference frame moving with speed $\V$
(namely, the rest frame of the travelling wave). 
Existence of three functionally-independent first integrals \eqref{eq:firstintegral} will then allow the traveling wave ODE \eqref{eq:4thorderODE} to be 
reduced to a first-order ODE, $f'=I(f)$. 

Second, 
suppose a conservation law $D_t T+D_x X=0$ has vanishing flux 
at the endpoints of a spatial domain $\tilde\Omega$ in $x$
for a travelling wave solution $u=f(\xi)$ of the wave equation $u_{tt}=F$. 
Then the conservation law yields a conserved quantity 
\begin{equation}
\frac{d}{dt}\int_{\tilde\Omega} T\big|_{u=f(\xi)} \,dx 
= \frac{d}{dt}\int_{\Omega} T\big|_{u=f(\xi)} \,d\xi 
= 0 
\end{equation}
where the domain will be $\Omega = (-\infty,\infty)$ for solitary wave solutions, 
or $\Omega = (-L/2,L/2)$ for nonlinear periodic wave solutions with wavelength $L>0$. 
If the conserved density $T$ does not contain the variables $t$ and $x$ explicitly,
then $T|_{u=f(\xi)}$ will be an expression involving only $f$ and its derivatives up to fourth order (\ie/ $\xi$ will not appear explicitly).
We can simplify this expression by using the reduced ODE $f'=I(f)$ to eliminate all derivatives of $f$, 
so that $T|_{u=f(\xi)} = \tilde T(f)$. 
If the wave profile of a travelling wave solution $u=f(\xi)$ on the domain $\Omega$ 
has a single maximum with an amplitude $f_{\max}$ at the center $\xi=0$ of the domain 
and a minimum with an amplitude $f_{\min}$ at the domain endpoints $\xi=\partial\Omega$, 
then the conserved quantity for this solution can be expressed as 
\begin{equation}\label{eq:conservedquantity}
\int_{\Omega} T\big|_{u=f(\xi)}\,d\xi
= 2\int_{f_{\min}}^{f_{\max}} \frac{\tilde T(f)}{I(f)}df 
\end{equation}
assuming that the wave profile is symmetric around $\xi=0$. 
This expression can be evaluated without the need for the explicit form of the solution $f(\xi)$.

\subsection{Highly nonlinear wave equation}

We will now apply the method outlined in section~\ref{sec:intro} 
to derive all travelling wave solutions 
for the highly nonlinear wave equation \eqref{eq:tilu-pde}. 

To begin, 
we substitute the traveling wave expression 
\begin{equation} 
\u = \f(\tilde\xi), 
\quad
\tilde\xi = x-\V t, 
\quad
\V =\const
\end{equation} 
into the wave equation \eqref{eq:tilu-pde}. 
This yields the fourth-order ODE
\begin{equation}\label{eq:f-ode}
(\V/c)^2 \f'' = (\f')^{k-1} \f'' + \alpha (\f')^{k-3}(\f'')^3 + \beta (\f')^{k-2}\f''\f''' +\gamma (\f')^{k-1} \f''''
\end{equation}
where $c$, $\alpha$, $\beta$, $\gamma$ are given by expressions \eqref{eq:tilu-pde-c}--\eqref{eq:tilu-pde-coeffs},
and where the nonlinearity exponent $k$ obeys
\cite{Spence1968,Johnson1985,Johnson2005}
\begin{equation}\label{eq:k-range}
k>1 .
\end{equation}

The physical variable that will physically describe a propagating wave 
is the strain $v = \u_x =-u_x >0$
(which is positive because the underlying discrete dynamical system has no restoring forces). 
Hence, we require the condition
\begin{equation}\label{eq:f-cond}
\f'>0
\end{equation}
on the travelling wave profile $\f(\tilde\xi)$. 
Moreover, 
since only derivatives of $\f(\tilde\xi)$ appear in the fourth-order ODE \eqref{eq:f-ode}, 
we will rewrite it in terms of the strain variable 
\begin{equation}\label{eq:g}
v=\g=\f'>0
\end{equation}
which gives a third-order ODE
\begin{equation}\label{eq:g-ode}
(\V/c)^2 \g' = 
\g^{k-1} \g' + \alpha \g^{k-3} (\g')^3 + \beta \g^{k-2}\g'\g'' +\gamma \g^{k-1} \g''' . 
\end{equation}

We now seek to find all solitary wave solutions and all periodic wave solutions 
of the strain ODE \eqref{eq:g-ode}. 

The first step in our method is to make use of the conservation laws \eqref{eq:TX1}--\eqref{eq:TX4}
derived in section~\ref{sec:prelims}
for the highly nonlinear wave equation \eqref{eq:tilu-pde}. 
By applying the direct reduction approach \eqref{eq:firstintegral}, 
we obtain first integrals for the ODE \eqref{eq:g-ode}
from the conservation laws \eqref{eq:TX1}, \eqref{eq:TX3}, and \eqref{eq:TX4}. 
These three conservation laws, 
which do not contain $t$ and $x$ explicitly,  
respectively describe the mean longitudinal-momentum, the total momentum, and the total energy 
for solutions of the wave equation \eqref{eq:tilu-pde}. 

The first integral arising from the mean longitudinal-momentum \eqref{eq:TX1} 
is given by 
\begin{equation}
\C_1 =  
(\V/c)^2 \g -\tfrac{1}{k}\g^{k} -\tfrac{k-1}{6}R^2 \g^{k-2}\g'{}^2 -\tfrac{1}{3}R^2 \g^{k-1}\g'' 
=\const. 
\label{eq:g-C1}
\end{equation}
Physically, it measures the spatial flux of mean longitudinal-momentum 
in the rest frame of the travelling wave. 
Both the total momentum  \eqref{eq:TX3} and the total energy \eqref{eq:TX4}
yield the same first integral (up to a multiplicative constant)
\begin{equation}
\C_2 = 
\tfrac{1}{2}(\V/c)^2 \g^{2} - \tfrac{1}{k+1} \g^{k+1} - \tfrac{k-2}{6}R^2 \g^{k-1}\g'{}^2  - \tfrac{1}{3}R^2 \g^{k}\g'' 
=\const. 
\label{eq:g-C2}
\end{equation}
Physically, this first integral measures the spatial flux of total energy/momentum 
in the rest frame of the travelling wave. 
We use the first integral \eqref{eq:g-C1} to solve for $\g''$ 
in terms of $\g'$ and $\g$, 
and substitute this expression into the first integral \eqref{eq:g-C2}. 
This immediately yields the first-order separable ODE
\begin{equation}
(\g')^2 =  
(6/R^2) \g^{1-k} \big( \tfrac{1}{2}(\V/c)^2 \g^{2} - \C_1 \g  +  \C_2 - \tfrac{1}{k(k+1)} \g^{1+k} \big) . 
\label{eq:ode}
\end{equation}
We write this ODE in dimensionless form 
by introducing the scaling 
\begin{gather}
\g = \lambda g,
\quad
\tilde\xi = \omega\xi, 
\label{eq:scal-g-xi}\\
C_1 = -\C_1 k(k+1) \lambda^{-k}, 
\quad
C_2 = \C_2 k(k+1) \lambda^{-1-k}. 
\end{gather}
If we choose
\begin{equation}\label{eq:scaling}
\lambda = \left(\tfrac{1}{2} (\V/c)^2 k(k+1)\right)^{1/(k-1)} ,
\quad
\omega = \sqrt{\tfrac{1}{6}k(k+1)}R , 
\end{equation}
then the ODE becomes
\begin{equation}\label{eq:scal-ode}
(g')^2 =  G(g),
\end{equation}
with 
\begin{equation}\label{eq:scal-G}
G(g) = g^{1-k} (g^2 + C_1 g +C_2 - g^{1+k}), 
\quad
g>0
\end{equation}
where $\xi$ and $g(\xi)$ are now dimensionless,
and where $C_1$ and $C_2$ are dimensionless arbitrary constants. 

The next step in our method is to classify all solutions of the first-order separable ODE \eqref{eq:scal-ode}
by employing a modified energy analysis. 
We first write $C_2 = E$, 
where $E$ represents a dimensionless energy for a solution $g(\xi)$. 
We view $C_1$ as a free parameter on which the solutions $g(\xi)$ will depend. 
Then we introduce the potential energy function 
\begin{equation}\label{eq:V}
V(g) =  g^{1+k} -g(g+C_1), 
\end{equation}
and write the ODE in the form 
\begin{equation}\label{eq:ode_energy_form}
g^{k-1} g'{}^2 + V(g) = E 
\end{equation}
which is a sum of a kinetic-type energy term and a potential energy term. 
The main part of the analysis is that we need to understand 
the shape of the potential \eqref{eq:V} as a function of $g>0$. 
To begin, we note its first and second derivatives are given by:
\begin{equation}
V'(g) = (1 + k) g^{k} - 2g - C_1, 
\qquad 
V''(g) = k(1+k)g^{k-1} - 2.
\end{equation}
We then see that there is only one point at which the convexity of $V(g)$ changes:
\begin{equation}\label{eq:gstar}
g^* = \big( \tfrac{2}{k(k+1)}\big)^{1/(k-1)} . 
\end{equation}
Since $k$ belongs to the range \eqref{eq:k-range}, 
note that we have
\begin{equation}
0<g^*<1 . 
\end{equation}

We now find all of the critical points $g=g^\c$ of $V(g)$. 
The number and nature of these points will depend on the value of $C_1$. 
First, we observe that finding the roots of $V'(g)=0$ is equivalent to 
finding the intersection points of the curve $(1 + k) g^{k}$ and the line $2g +C_1$. 
Now consider the slope of the curve, which is given by $(1+k)k g^{k-1}$. 
At $g=0$, the slope vanishes. 
For $g>0$, the slope is increasing and becomes equal to the slope of the line 
at $g=g^*$. 
For $g>g^*$, the curve has a greater slope than the slope of the line. 
Hence, if $C_1\geq 0$, then for $g>0$ 
the curve will intersect the line at exactly one point $g=g^\c_1$, 
which will lie to the right of $g^*$. 
Since $V''(g^\c_1)>0$, 
we conclude that $g^\c_1$ is a local minimum of $V(g)$. 
Next, if $C_1<0$, then there are three cases, 
depending on the sign of the difference 
$(1 + k) g^*{}^{k} - (2g^* +C_1) = C_1^*-C_1$
where 
\begin{equation}\label{eq:C1star}
C_1^* = (1 + k) g^*{}^{k} - 2g^* = -\tfrac{2(k-1)}{k}g^* <0 . 
\end{equation}
If the sign of $C_1^*-C_1$ is negative, then for $g>0$ 
the line will intersect the curve in two distinct points 
$g=g^\c_1$ and $g=g^\c_2$ such that $0 < g^\c_1 < g^*<g^\c_2$. 
The critical point at $g^\c_1$ is a local maximum of $V(g)$,
while the other critical point $g^\c_2$ is a local minimum of $V(g)$. 
If $C_1^*-C_1$ is zero, then for $g>0$ 
the line will intersect the curve only at the point (\ie/, as a tangent line)
$g=g^*$,
which is an inflection point of $V(g)$. 
If the sign of $C_1^*-C_1$ is positive, then for $g>0$ 
the line will not intersect the curve. 
This completes the determination of all of the critical points of $V(g)$. 

\begin{figure}[!h]
\centering
\includegraphics[width=\textwidth]{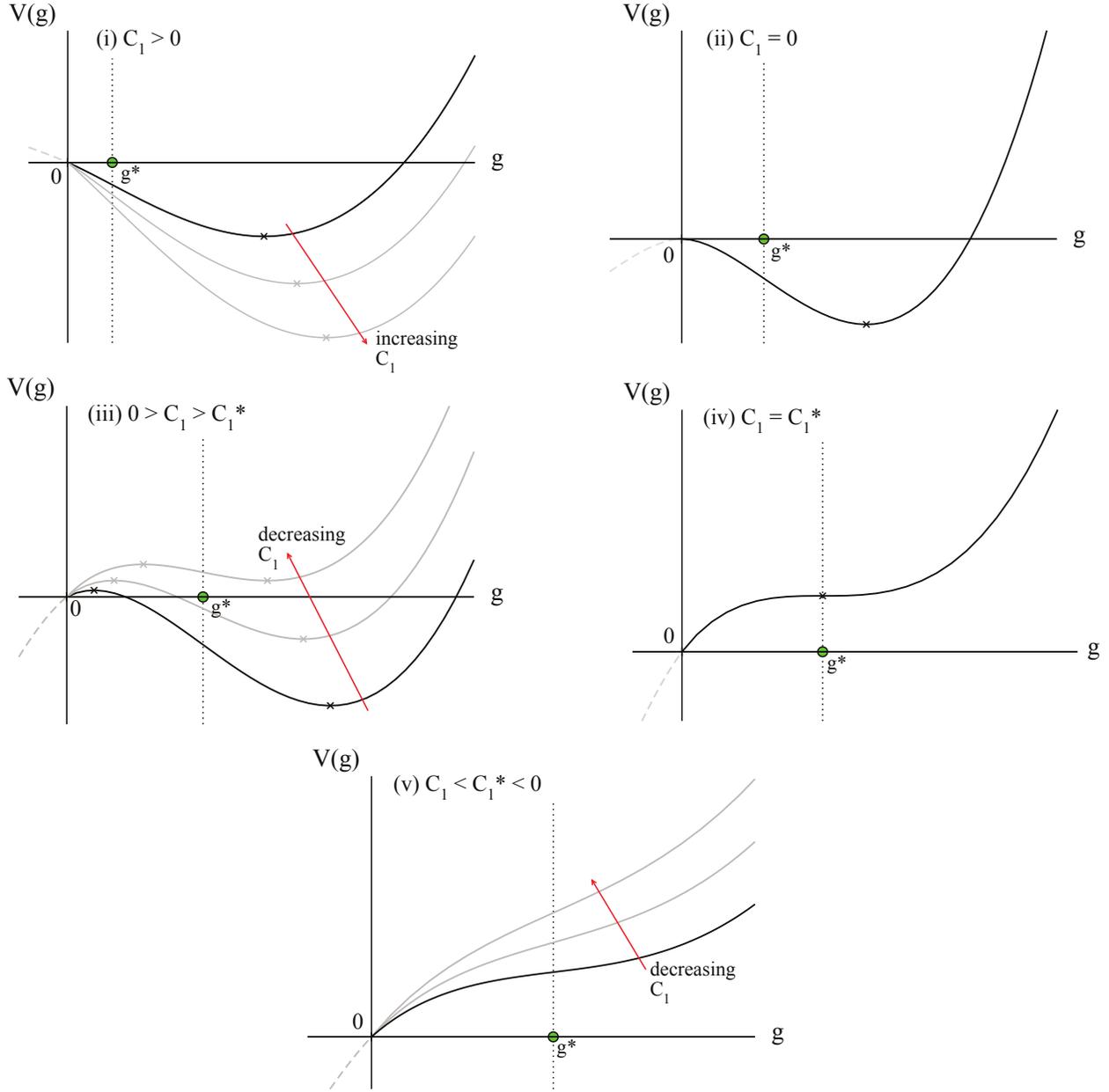}
\caption{Potential energy function $V(g) =  g^{1+k} -g(g+C_1)$ for the Hertz potential $k =3/2$. }
\label{fig:Vshape}
\end{figure}

Finally, we look at the asymptotic behaviour of $V(g)$ as $g\to 0$ and $g\to \infty$. 
For large $g$, the dominant term in $V(g)$ is $g^{1+k}\gg g(g+C_1)$, 
so then $V(g)\to \infty$. 
For $g$ near $0$, the dominant term in $V(g)$ is 
either $|C_1|g \gg g^{1+k}$ if $C_1\neq 0$,
or $g^2 \gg g^{1+k}$ if $C_1=0$. 
Combining this asymptotic analysis with the critical point analysis, 
we obtain a full picture of the potential energy function $V(g)$. 
In particular, there are five distinct cases, 
depending on the value of $C_1$, 
as illustrated in Fig.~\ref{fig:Vshape}. 

We can now classify the different types of solutions of ODE \eqref{eq:ode_energy_form}
by using an energy analysis based on the features of the energy potential $V(g)$. 
Consider the intersection of an energy line $E=\const$ and the potential $V(g)$ 
for $g>0$. 
The ODE \eqref{eq:ode_energy_form} shows that $g^{k-1} g'{}^2=0$
holds at the intersection point, 
and this implies $g'=0$ (since $g>0$). 
Hence, all such intersection points are critical points of the wave profile $g(\xi)$. 
This leads to four different types of solutions, 
two of which have the feature that $g(\xi)$ vanishes at isolated points $\xi$. 

Additional types of solutions arise if we also consider intersection points given by $g=0$, 
which can occur (only) if $E=0$. 
Compared to the four main types, 
these additional solutions exhibit different behaviour at the points $\xi$ where $g(\xi)=0$. 

A summary of all of the different solution types is stated in Table~\ref{table:types}. 
Details will be derived in subsequent subsections \ref{periodic}--\ref{limiting}. 

\begin{table}[htb]
\centering
\caption{Types of travelling wave solutions}
\label{table:types}
\begin{tabular}{c|c|c|c|c}
\hline
Intersection points & Type of point & Nonlinearity & Solution type & Domain 
\\
$V(g) = E$ & $g$ & $k$ & $g(\xi)$ & $\xi$
\\
\hline
\hline
two $g>0$
& 
one local maximum
&
$>1$
&
solitary wave
& 
$(-\infty, \infty)$
\\
&
one non-critical
&
&
\\
\hline
one $g>0$
& 
local maximum
& 
$>1$
&
cusp-nodal 
&
$(-\infty, \infty)$
\\
&
&
&
solitary wave
&
\\
\hline
two $g>0$
&
non-critical
&
$>1$
&
periodic wave
& 
$[\tfrac{-L}{\,2}, \tfrac{L}{2}]$
\\
\hline
one $g>0$
&
non-critical 
&
$>1$
&
cusp-nodal
&
$[\tfrac{-L}{\,2}, \tfrac{L}{2}]$
\\
&
&
&
periodic wave
&
\\
\hline
\hline
one $g=0$ 
&
non-critical
&
$<2$
&
nodal periodic wave
&
$[\tfrac{-L}{\,2}, \tfrac{L}{2}]$
\\
one $g >1$
&
non-critical
&
&
&
\\
\hline
one $g=0$ 
&
non-critical
&
$=2$
&
corner-nodal 
&
$[\tfrac{-L}{\,2}, \tfrac{L}{2}]$
\\
one $g >1$
&
non-critical
&
&
periodic wave
&
\\
\hline
one $g=0$ 
&
non-critical
&
$>2$
&
cusp-nodal 
&
$[\tfrac{-L}{\,2}, \tfrac{L}{2}]$
\\
one $g >1$
&
non-critical
&
&
periodic wave
&
\\
\hline
one $g=0$ 
&
critical
&
$<3$
&
nodal periodic wave
& 
$[\tfrac{-L}{\,2}, \tfrac{L}{2}]$
\\
one $g=1$ 
&
non-critical
&
&
&
\\
\hline
one $g=0$ 
&
critical
&
$=3$
&
corner-nodal 
& 
$[\tfrac{-L}{\,2}, \tfrac{L}{2}]$
\\
one $g=1$ 
&
non-critical
&
&
periodic wave
&
\\
\hline
one $g=0$ 
&
critical
&
$>3$
&
cusp-nodal 
&
$[\tfrac{-L}{\,2}, \tfrac{L}{2}]$
\\
one $g=1$ 
&
non-critical
&
&
periodic wave
&
\\
\hline
\hline
\end{tabular}
\end{table}

The presence of a node in a travelling wave $g(\xi)$ 
means that the strain \eqref{eq:g} is no longer positive 
but vanishes at isolated points $\xi_0$. 
This can be viewed as a limiting case of vanishingly small strain. 

To understand how the different types of solution arise, 
we integrate the first-order separable ODE \eqref{eq:ode_energy_form} to get 
\begin{equation}\label{eq:ode_integral}
\pm \int_{g_0}^{g} \frac{g^{ (k-1)/2 }} {\sqrt{E - V(g)}} \,dg = \xi 
\end{equation}
where $g_0>0$ is a root of $V(g)=E$. 
Then we can factorize the denominator expression 
\begin{equation}\label{eq:ode_factored}
E -V(g) = (g-g_0)^m A(g)
\end{equation}
where $m$ is some positive integer (\ie/, the degeneracy of the root)
and $A(g)$ is some function of $g$ such that $A(g_0)$ is non-singular. 
This factorization will hold under the conditions that 
$E-V(g)$ and its derivatives up to order $m-1$ vanish at $g=g_0$. 
Note, from the shape of $V(g)$, 
the maximum possible degeneracy is $m=3$,
and the root $g_0$ is a critical point of $V(g)$ iff $m\neq1$. 
Next we expand the integral \eqref{eq:ode_integral} 
asymptotically for $g$ near $g_0>0$:
\begin{equation}\label{eq:g-near-g0}
\begin{aligned}
\int \frac{g^{ (k-1)/2 }} {\sqrt{E - V(g)}} \,dg 
& \sim \dfrac{g_0^{(k-1)/2}}{\sqrt{|A(g_0)|}} \int\frac{dg}{|g-g_0|^{m/2}} 
\\
& \sim \dfrac{g_0^{(k-1)/2}}{\sqrt{|A(g_0)|}} 
\begin{cases}
2|g-g_0|^{1/2} ,
& m=1\\
\ln|g-g_0|, 
& m=2\\
-2|g-g_0|^{-1/2} ,
& m=3 . 
\end{cases}
\end{aligned}
\end{equation}
In the case $m=1$, this integral converges, 
and so $g(\xi)$ will equal $g_0$ at a finite value $\xi=\xi_0$. 
In the other two cases, the integral diverges, 
whereby $g(\xi)$ will asymptotically approach $g_0$ as $|\xi|$ gets large. 
This asymptotic tail will have exponential decay of $|g-g_0|\to 0$ in the case $m=2$ 
and power decay of $|g-g_0|\to 0$ in the case $m=3$. 
The global behaviour of the solution $g(\xi)$ as a function of $\xi$ depends on whether 
the domain for the integral \eqref{eq:ode_integral} is given by 
$g\geq g_0$ or $g\leq g_0$, 
as determined by the condition $V(g)\leq E$ so that $\sqrt{E-V(g)}$ is well-defined. 

\begin{figure}[!h]
\centering
\includegraphics[width=\textwidth]{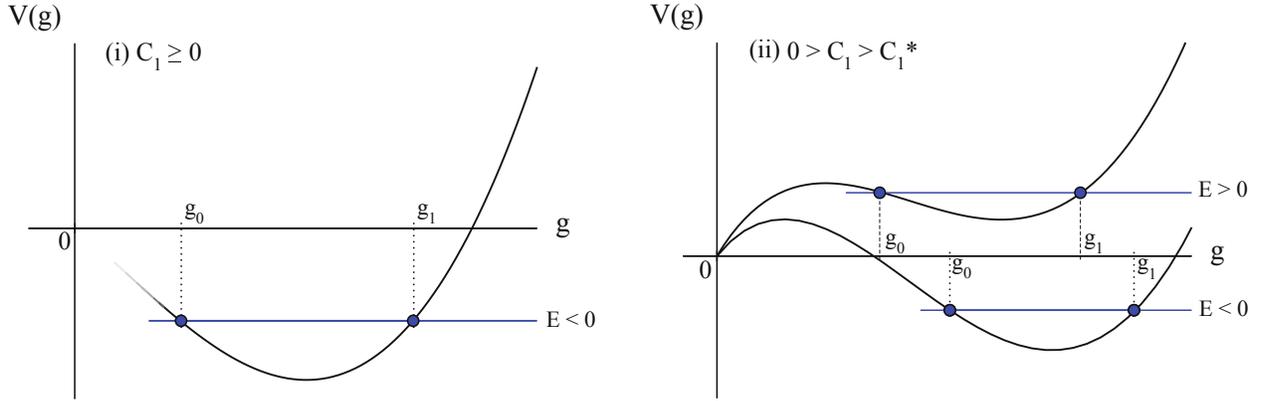}
\caption{Conditions for a periodic wave solution for the potential energy function $V(g) =  g^{1+k} -g(g+C_1)$ for the Hertz potential $k =3/2$. }
\label{fig:Case-1}
\end{figure}

\begin{figure}[!h]
\centering
\includegraphics[width=0.5\textwidth]{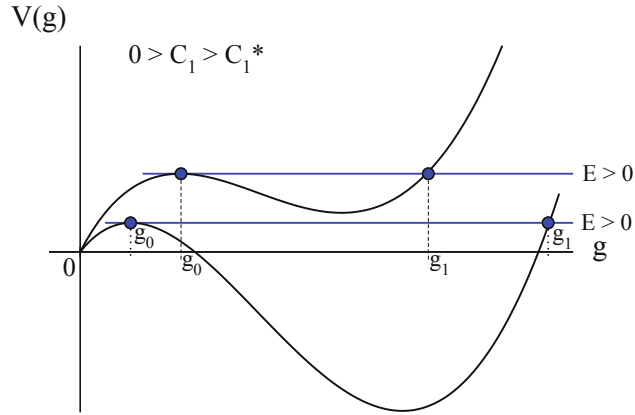}
\caption{Conditions for a solitary wave solution for the potential energy function $V(g) =  g^{1+k} -g(g+C_1)$ for the Hertz potential $k =3/2$. }
\label{fig:Case-1prime}
\end{figure}

When $V(g)\leq E$ holds for some $g\geq g_0>0$, 
the shape of $V(g)$ shows that $C_1$ must be greater than $C_1^*$,
and that $V(g)=E$ has a second root $g_1$. 
This root satisfies $g_1 >g_0>0$ and is not a critical point of $V(g)$. 
Consequently, 
the domain for the solution integral \eqref{eq:ode_integral} 
will be $g_0 \leq g\leq g_1$,
and the resulting solution $g(\xi)$ 
will have the properties that it will reach $g_1$ at a finite value of $\xi$, 
and that it will reach $g_0$ 
either at a different finite value of $\xi$ if $g_0$ is not a critical point of $V(g)$, see Fig.~\ref{fig:Case-1} 
or asymptotically for large $|\xi|$ if $g_0$ is a critical point of $V(g)$, see Fig.~\ref{fig:Case-1prime}. 
Moreover, in the latter case, the shape of $V(g)$ shows that $V''(g_0)\neq 0$,
so the degeneracy of $g_0$ will be $m=2$. 
These two possibilities for the global behaviour of $g(\xi)$ thereby describe, respectively,
a periodic wave (obtained by piecing together translated copies of the wave profile) 
whose maximum and mininum amplitudes are $g_1$ and $g_0$, 
and a solitary wave with a maximum amplitude $g_1$ and an exponential tail decaying to $g_0$. 

\begin{figure}[!h]
\centering
\includegraphics[width=\textwidth]{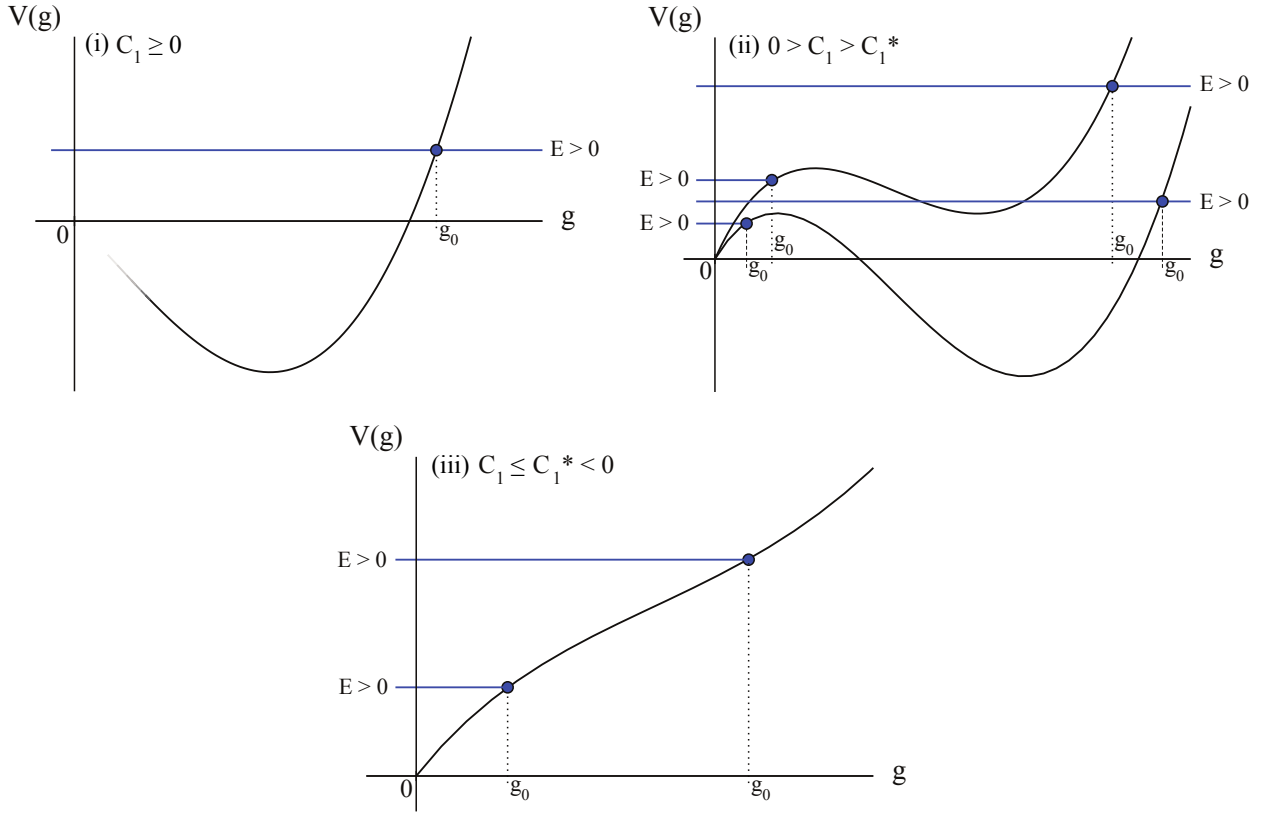}
\caption{Conditions for a cusped periodic wave solution for the potential energy function $V(g) =  g^{1+k} -g(g+C_1)$ for the Hertz potential $k =3/2$. }
\label{fig:Case-2}
\end{figure}

\begin{figure}[!h]
\centering
\includegraphics[width=\textwidth]{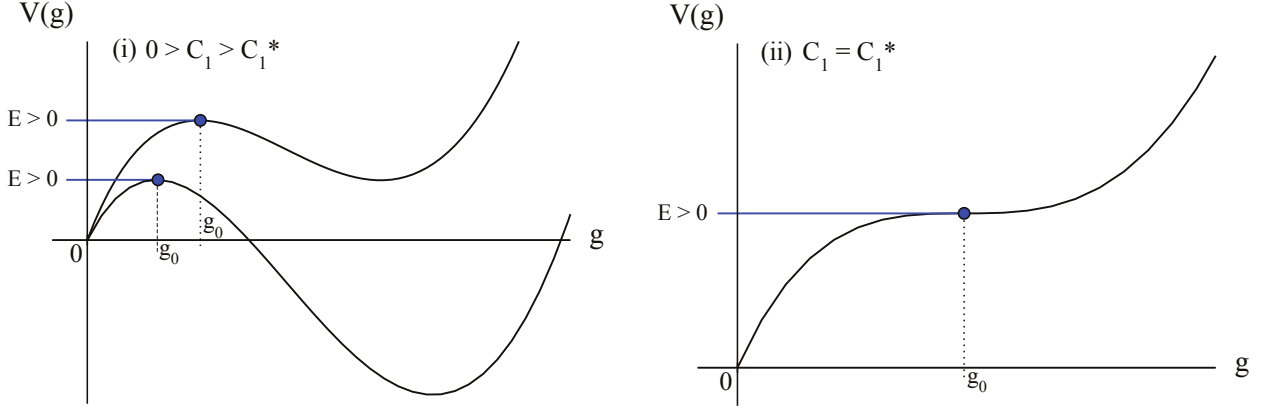}
\caption{Conditions for a cusped solitary wave solution for the potential energy function $V(g) =  g^{1+k} -g(g+C_1)$ for the Hertz potential $k=3/2$. }
\label{fig:Case-2prime}
\end{figure}

When $V(g)\leq E$ holds for some $g\leq g_0>0$ with $E>0$, 
the shape of $V(g)$ shows that $V(g)=E>0$ has no roots other than $g_0$
and that there is no restriction on $C_1$. 
Consequently, 
the domain for the solution integral \eqref{eq:ode_integral} 
will be $0\leq g\leq g_0$. 
The resulting solution $g(\xi)$ will reach $0$ at a value of $\xi$ 
determined by the convergence of the integral \eqref{eq:ode_integral} near $g=0$. 
In particular, due to $k>1$ and $E>0$, 
we have 
\begin{equation}
\int \frac{g^{ (k-1)/2 }} {\sqrt{E - V(g)}} \,dg 
\sim \dfrac{1}{\sqrt{E}} \int g^{ (k-1)/2 }\,dg =0
\end{equation}
and 
\begin{equation}
(g')^2 \sim E g^{1-k} =\infty . 
\end{equation}
Hence, $g(\xi)$ will go to $0$ at a finite value of $\xi$, while $g'(\xi)$ will blow up. 
This describes a nonlinear wave $g(\xi)$ that has a nodal cusp. 
Furthermore, this solution will reach $g_0$ 
either at a finite value of $\xi$ if $g_0$ is a not critical point of $V(g)$, see Fig.~\ref{fig:Case-2}, 
or asymptotically for large $|\xi|$ if $g_0$ is a critical point of $V(g)$, see Fig.~\ref{fig:Case-2prime}.
These respective cases describe a cusped periodic wave 
(obtained by piecing together translated copies of the wave profile), 
and a cusped solitary wave. 
In the latter case, 
the shape of $V(g)$ shows that either $V''(g_0)\neq 0$ if $C_1>C_1^*$,
whereby the degeneracy of $g_0$ will be $m=2$,  
or $V''(g_0)= 0$ if $C_1=C_1^*$,
whereby the degeneracy of $g_0$ will be $m=3$. 
These two possibilities correspond to the tail of the cusped solitary wave 
having either exponential decay or power decay of $|g-g_0|\to 0$ for large $|\xi|$. 

\begin{figure}[!h]
\centering
\includegraphics[width=\textwidth]{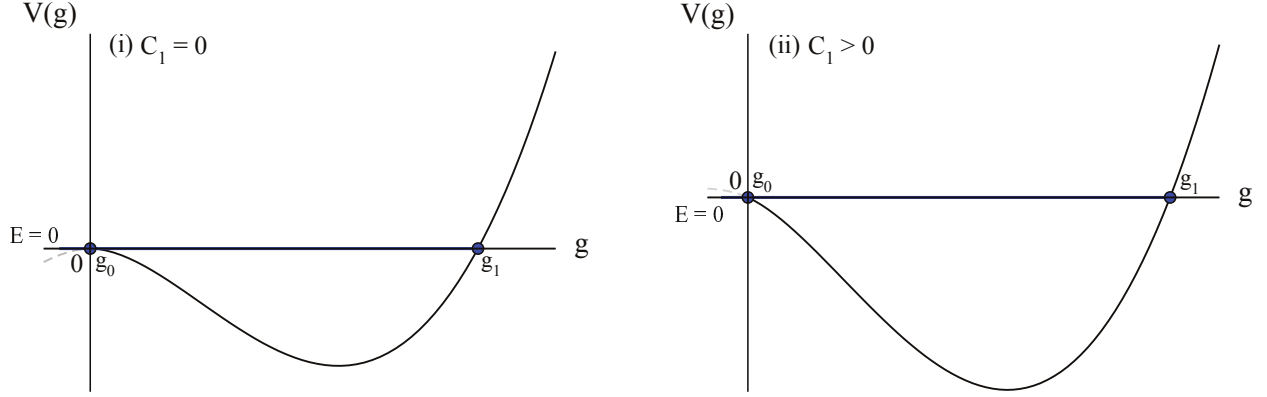}
\caption{Conditions for the limiting case $E=0$ (nodal, corner, cusp) periodic wave solutions for the potential energy function $V(g) =  g^{1+k} -g(g+C_1)$ for the Hertz potential $k=3/2$. }
\label{fig:Cases34}
\end{figure}

A limiting case occurs when $V(g)\leq E=0$ holds for some $g\geq g_0=0$,
where $V(0)=0$ and $V'(0)=-C_1$. 
In this case, as shown by the shape of $V(g)$, see Fig.~\ref{fig:Cases34},
we have $C_1 \geq 0$, 
and $V(g)=E=0$ has a second root $g_1>g^*>0$
which is not a critical point of $V(g)$. 
Moreover, we see $g_0=0$ is a critical point of $V(g)$ iff $C_1=0$. 
The domain for the solution integral \eqref{eq:ode_integral} 
thereby will be $0\leq g\leq g_1$,
and the resulting solution $g(\xi)$ will reach $g_1>0$ at a finite value of $\xi$. 
The solution $g(\xi)$ will reach $g_0=0$ at a value of $\xi$ 
determined by expanding the solution integral \eqref{eq:ode_integral} 
asymptotically for $g$ near $0$: 
\begin{equation}
\begin{aligned}
\int_{0}^{g} \frac{g^{ (k-1)/2 }} {\sqrt{E - V(g)}} \,dg 
= \int_{0}^{g}  \frac{g^{k/2-1}}{\sqrt{C_1+g- g^{k}}} \,dg 
& \sim \begin{cases}
\tfrac{2}{k-1} g^{(k-1)/2} \sim 0 , 
& C_1=0\\
\tfrac{2}{k\sqrt{C_1}} g^{k/2} \sim 0 , 
& C_1 >0
\end{cases}
\\& 
\sim 0
\end{aligned}
\end{equation}
which is due to $k>1$. 
Thus, since this integral converges to yield $0$, 
we can express the solution $g(\xi)$ as 
\begin{equation}\label{eq:g0=0}
\pm \int_{0}^{g}  \frac{g^{k/2-1}}{\sqrt{C_1+g- g^{k}}} \,dg = \xi 
\end{equation}
where $g(0)=0$. 
The ODE \eqref{eq:ode_energy_form} then shows that 
\begin{equation}
(g')^2 
\sim 
\begin{cases}
g^{3-k}, & C_1 =0
\\
C_1g^{2-k}, & C_1 >0
\end{cases}
\quad\text{ as } g\to 0. 
\end{equation}
This implies 
\begin{equation}
|g'(0)| = 
\begin{cases}
0, & k<3
\\
1, & k=3
\\
\infty, & k>3
\end{cases}
\quad\text{ when } V'(0) =0, ~\text{Fig.~\ref{fig:Cases34}(i)},
\end{equation}
and 
\begin{equation}
|g'(0)| = 
\begin{cases}
0, & k<2
\\
V'(0), & k=2
\\
\infty, & k>2
\end{cases}
\quad\text{ when } V'(0) < 0, ~\text{Fig.~\ref{fig:Cases34}(ii)}.
\end{equation}
Therefore, 
the solution $g(\xi)$ describes a periodic wave 
in which $\xi=0$ modulo $L$ is a node that exhibits a cusp, a corner, or a local minimum,
depending on the values of $k$ and $V'(0)=-C_1$,
where $L$ is the wavelength. 
(Note that the case $V'(0)=0$ distinguishes when the set of constant strain solutions, 
$g=\const$, includes $g=0$.) 

We remark that the presence of a node in a travelling wave $g(\xi)$ 
means that the strain \eqref{eq:g} is no longer positive 
but vanishes at isolated points $\xi_0$. 
This can be viewed as a limiting case of vanishingly small strain. 
We also remark that the wave profile $g(\xi)$ for cusped waves and peaked waves 
is not a differentiable function of $\xi$ at the locations of the cusps and peaks. 
Thus, at these values of $\xi$, 
$g(\xi)$ does not satisfy the ODE \eqref{eq:ode_energy_form}. 
As a consequence, for $g(\xi)$ to make sense as an actual solution, 
we would need to seek a suitable weak formulation of the ODE, 
similarly to what is done for peakon wave solutions 
\cite{Con-Str,Con-Mol-2000,AncoRecio}
of the Camassa-Holm equation and other related nonlinear wave equations. 

In the next step of our method, 
based on the preceding analysis of the solution integral \eqref{eq:ode_integral}, 
we determine the conditions on the constants $E$, $g_0$, and $C_1=V'(0)$
under which the various types of solutions $g(\xi)$ 
(listed in Table~\ref{table:types}) 
are produced.  
For carrying out this analysis, 
it will be useful to introduce the notation
\begin{equation}\label{eq:S}
S_n(a, b) = S_n(b,a) = \frac{a^n-b^n}{a-b} ,
\quad
a\neq b,
\quad
n>0 . 
\end{equation}
As a function of $a$ and $b$, 
$S_n(a,b)$ has the following properties that we will need. 
First, for $a>b>0$ and $b>a>0$, 
$S_n(a,b)$ is non-singular when $n\geq 0$; 
$\partial_a S_n(a,b)$ and $\partial_a^2 S_n(a,b)$ are non-singular when $n\geq 1$ and $n\geq 2$, respectively. 
Second, $S_n(a,b)$ obeys the inequalities
\begin{equation}\label{eq:S-upper}\\
S_n(a, b) \leq n\frac{a^n+b^n}{a+b},
\quad n\geq 0
\end{equation}
and
\begin{align}
S_n(a, b) >S_n(a,c) >0, 
\quad
n>0
\label{eq:S-property1}\\
\partial_a S_n(a, b) >\partial_a S_n(a, c) >0, 
\quad 
n>1
\label{eq:S-property2}\\
\partial_cS_n(c, a) >\partial_c S_n(c,b) >0, 
\quad 
n>1
\label{eq:S-property2'}\\
\partial_a^2 S_n(a, b) >\partial_a^2 S_n(a, c) >0,
\quad
n>2
\label{eq:S-property3}
\end{align}
for $a>b>c\geq 0$. 
Third, $S_n(a,b)$ has the limits
\begin{equation}\label{eq:S-limit}
S_n(a, a) = n a^{n-1}, 
\quad
\partial_a S_n(a, b)|_{b=a} = \tfrac{1}{2}n(n-1)a^{n-2}
\end{equation}
for $a>0$.

\subsection{Periodic waves}\label{periodic}

To obtain a periodic nonlinear wave solution $g(\xi)>0$, 
we must have 
\begin{align}
&\begin{aligned}
& C_1\geq 0:\quad 0>E>V(g_1^\c) 
\\
& 0>C_1> C_1^*:\quad V(g^\c_1)>E>V(g^\c_2) 
\end{aligned}
\label{eq:periodic-1}
\\
& E=V(g_0)=V(g_1),
\quad
0<g_0<g_1
\label{eq:periodic-2}
\\
& V'(g_0) < 0, 
\quad
V'(g_1) >0 
\label{eq:periodic-3}
\\
& V(g) \leq V(g_0) \text{ for } g_0\leq g\leq g_1
\label{eq:periodic-4}
\end{align}
where $V'(g_1^\c) = V'(g_2^\c) = 0$. 

The second condition \eqref{eq:periodic-2}
yields $E= g_0^{k+1} - g_0(g_0 + C_1) = g_1^{k+1} - g_1(g_1 + C_1)$. 
These two equations determine 
\begin{align}
C_1 & = S_{k+1}(g_1,g_0)- (g_1+ g_0)
\\
E & = g_1g_0\big(1 - S_{k}(g_1,g_0)\big)
\label{eq:periodic_E}
\end{align}
in terms of $g_0$ and $g_1$. 
Then we have 
\begin{equation}\label{eq:periodic_V}
V(g) = g\big( g^k -g +g_1+ g_0 - S_{k+1}(g_1,g_0) \big)
\end{equation}
which gives 
\begin{equation}
E-V(g) = (g_1-g)(g-g_0)A(g)
\end{equation}
where
\begin{equation}\label{eq:periodic_A}
A(g) = \frac{S_{k+1}(g,g_1)-S_{k+1}(g,g_0)}{g_1-g_0} -1
= \frac{g_1S_{k}(g,g_1)-g_0S_{k}(g,g_0)}{g_1-g_0}  -1 . 
\end{equation}
Next, we find that the third condition \eqref{eq:periodic-3} yields
\begin{equation}\label{eq:periodic-Sk}
g_1(S_{k}(g_1,g_0) -1) >  g_0(kg_0^{k-1}-1)
\end{equation}
and 
\begin{equation}
g_1(kg_1^{k-1}-1) > g_0(S_{k}(g_1,g_0) -1) . 
\end{equation}
Comparing these two inequalities, 
we see that the second one follows from the first one, 
because $k(g_1^k + g_0^k) -(g_1+g_0)S_{k}(g_1,g_0) >0$ 
due to property \eqref{eq:S-upper}, 
using 
\begin{equation}\label{eq:periodic-g0g1}
0<g_0<g_1 . 
\end{equation}
Then, the remaining conditions \eqref{eq:periodic-4} and \eqref{eq:periodic-1} 
will hold because $V(g_0)-V(g)$ is increasing for both $g-g_0$ and $g_1-g$ increasing 
up to the single critical point $g=g_2^\c$ between $g_0$ and $g_1$.
Finally, we note that inequality \eqref{eq:periodic-Sk} imposes a simple lower bound 
on $g_1$ as follows. 
Consider the extended inequality $g_1(S_{k}(g_1,g_0) -1) \geq g_0(kg_0^{k-1}-1)$
with $0<g_0\leq g_1$. 
This inequality can be re-written as $B_0(g_0)+B_1(g_0)g_1 \geq C(g_1)$,
which says that the line $B_0(g_0)+B_1(g_0)g_1$ lies above the curve $C(g_1)$,
where $B_0(g)= g^2(kg^{k-1}-1)$, $B_1(g)=g(2-(k+1)g^{k-1})$, 
$C(g)=g^2(1-g^{k-1})$. 
We see that if $g_0=g_1$ then 
$B_0(g_1)+B_1(g_1)g_1 = C(g_1)$ and $B_1(g_1) = C'(g_1)$,
so thus the line touches the curve as a tangent line. 
Hence, the boundary of inequality \eqref{eq:periodic-Sk} 
consists of the points $g_1>0$ at which 
the line $B_0(g_0)+B_1(g_0)g_1$ is tangent to the curve $C(g_1)$. 
The minimum $g_1$ in this set of tangent points can be seen geometrically to come from 
the critical point $g_0^*$ of the set of lines $B_0(g_0)+B_1(g_0)g_1$
as given by $B_0'(g_0^*)=0$ and $B_1'(g_0^*)=0$. 
We have $B_0'(g)= k(k+1)g^{k} - 2g$ and $B_1'(g)=2-k(k+1)g^{k-1}$, 
which yields $g_0^*= g^*$,
where $g^*$ is expression \eqref{eq:gstar}. 
This gives $g_1= g^*$ for the minimum of $g_1$ in the set of tangent points. 
We thus have the lower bound 
\begin{equation}\label{eq:periodic-g1gstar}
g_1 > g^*
\end{equation}
under which the inequality \eqref{eq:periodic-Sk} will hold. 

Therefore, a periodic wave exists iff the constants $g_0$ and $g_1$
satisfy the inequalities \eqref{eq:periodic-Sk}, \eqref{eq:periodic-g0g1}, \eqref{eq:periodic-g1gstar}. 
Then the solution integral \eqref{eq:ode_integral} is given by 
\begin{equation}\label{eq:periodic_ode_integral}
\pm \int_{g_0}^{g} \frac{g^{ (k-1)/2 }}{\sqrt{(g_1-g)(g-g_0)A(g)}} \,dg = \xi \mp L/2
\end{equation}
with $g_0\leq g\leq g_1$,
where 
\begin{equation}
L = 2\int_{g_0}^{g_1} \frac{g^{ (k-1)/2 }}{\sqrt{(g_1-g)(g-g_0)A(g)}} \,dg >0
\end{equation}
is the wavelength. 
The wave profile $g(\xi)$ for this solution \eqref{eq:periodic_ode_integral}
will reach a local maximum $g_1$ at $\xi=0$ modulo $L$
and a local minimum $g_0$ at $\xi=\pm L/2$ modulo $L$.

\subsection{Cusped periodic waves}

This type of periodic solution $g(\xi)\geq 0$ arises when 
\begin{align}
& E >0
\label{eq:cusp-1}
\\
& E=V(g_0) 
\label{eq:cusp-2}
\\
& V'(g_0) > 0
\label{eq:cusp-3}
\\
& V(g) \leq V(g_0) \text{ for } 0<g\leq g_0 . 
\label{eq:cusp-4}
\end{align}

The second condition \eqref{eq:cusp-2}
yields $E= g_0^{k+1} - g_0(g_0 + C_1)$,
which determines
\begin{equation}
C_1 = g_0^{k} - g_0 - E/g_0 . 
\end{equation}
This gives
\begin{equation}\label{eq:cusped_V}
V(g) = (g^{k}-g_0^k -g+g_0 + E/g_0)g 
\end{equation}
and thus we have 
\begin{equation}
E-V(g) = (g_0-g)A(g)
\end{equation}
where 
\begin{equation}\label{eq:cusped_A}
A(g) = S_{k}(g,g_0)g - g +  E/g_0 = S_{k+1}(g,g_0) - g -g_0^k +  E/g_0 . 
\end{equation}
From the first and third and conditions \eqref{eq:cusp-1} and \eqref{eq:cusp-3},
we have 
\begin{align}
E/g_0 & > 0, 
\label{eq:cusp-E-1}
\\
E/g_0 & > g_0(1 - kg_0^{k-1}).
\label{eq:cusp-E-2}
\end{align}
The fourth condition \eqref{eq:cusp-4} reduces to $A(g)\geq 0$. 
We note $A(0)=E/g_0>0$ and $A(g_0)=k g_0^k - g_0 +  E/g_0>0$
by using inequalities \eqref{eq:cusp-E-1}--\eqref{eq:cusp-E-2} 
and property \eqref{eq:S-limit}. 
Now consider $A'(g) = \partial_gS_{k+1}(g,g_0) -1$ from expression \eqref{eq:cusped_A}. 
Using properties \eqref{eq:S-property2} and \eqref{eq:S-limit}, 
we see $A'(g) \leq A'(g_0)=\partial_gS_{k+1}(g_0,g_0) -1 = (g_0/g^*)^{k-1} -1$
for $0<g\leq g_0$, 
from expressions \eqref{eq:cusped_A} and \eqref{eq:gstar}. 
Hence, we obtain $A'(g)\leq 0$ if $g_0\leq g^*$, 
which then implies $A(g)\geq A(g_0)$ for $0<g\leq g_0$. 
Since $A(g_0)>0$, 
we conclude $A(g)\geq 0$ for $0<g\leq g_0$ in the case $g_0\leq g^*$. 
The complementary case $g_0> g^*$ is more complicated 
and leads to two subcases as follows. 
We first note 
$A'(g) \geq \partial_gS_{k+1}(g,g) -1 = \tfrac{1}{2}(k+1)k g^{k-1} -1= (g/g^*)^{k-1} -1$
from properties \eqref{eq:S-property2} and \eqref{eq:S-limit}. 
This implies $A'(g)>0$ for $g>g^*$, 
whereby $A(g)$ is increasing for $g>g^*$. 
Also, we have $A''(g)= \partial_g^2S_{k+1}(g,g_0) >0$ 
by property \eqref{eq:S-property3}. 
Next we note that $A'(0)=  \partial_gS_{k+1}(g,g_0)|_{g=0} -1= g_0^{k-1}-1$ 
by direct evaluation. 
This gives a case split, depending on whether $g_0>1$ or $g_0<1$. 
If $g_0\geq1$, we have $A'(0)\geq 0$, 
which then implies that $A(g)$ is increasing for $0<g\leq g_0$,
since $A''(g)>0$. 
We thereby conclude that $A(g)>0$ due to $A(0)>0$. 
If instead $g_0<1$, then we have $A'(0)<0$, 
which implies $A(g)$ has a unique minimum for some $g=g_0^*\leq g^*$, 
since $A'(g)>0$ for $g>g^*$ 
and $A''(g)>0$ for $0<g<g_0$. 
This minimum is determined by 
\begin{equation}\label{eq:cusped-min-A}
\partial_gS_{k+1}(g,g_0)|_{g=g_0^*}=1 .
\end{equation}
Hence, we conclude $A(g)\geq A(g_0^*)$, 
and then we require $A(g_0^*)\geq 0$ 
so that $A(g)\geq 0$ will hold for $0<g\leq g_0$. 
This yields the condition $E/g_0 \geq (1-S_{k}(g_0^*,g_0))g_0^*$
from expression \eqref{eq:cusped_A}. 
Finally, 
we note that equation \eqref{eq:cusped-min-A} can be expressed in an equivalent form
by expanding out the derivative, which gives 
$S_{k+1}(g_0^*,g_0)=(k+1){g_0^*}^k -g_0^* +g_0$. 
In turn, 
we can express $S_{k+1}(g_0^*,g_0)={g_0^*}^k + g_0 S_{k}(g_0^*,g_0)$,
yielding 
\begin{equation}\label{eq:cusped-min-A-alt}
g_0(1- S_{k}(g_0^*,g_0)) = g_0^*(1-k{g_0^*}^{k-1}) .
\end{equation}

Therefore, 
this analysis shows that a cusped periodic wave exists 
iff the constants $g_0$ and $E$ satisfy the inequalities 
\begin{subequations}
\begin{align}
E & >0 , 
\quad
g_0\geq 1
\label{eq:cusp-E-g0-1}
\\
E & \geq (1-k{g_0^*}^{k-1}) {g_0^*}^2>0 , 
\quad
g^*<g_0 <1
\label{eq:cusp-E-g0-2}
\\
E & \geq  (1-kg_0^{k-1})g_0^2 >0, 
\quad
0<g_0 <g^*
\label{eq:cusp-E-g0-3}
\end{align}
\end{subequations}
where $g_0^*$ is given by equation \eqref{eq:cusped-min-A} 
(or, equivalently, equation \eqref{eq:cusped-min-A-alt}). 
The solution integral \eqref{eq:ode_integral} then can be expressed as 
\begin{equation}\label{eq:cuspedperiodic_ode_integral}
\pm \int_{g}^{g_0} \frac{g^{ (k-1)/2 }}{\sqrt{(g_0-g)A(g)}} \,dg = \xi
\end{equation}
with $0<g\leq g_0$. 
This solution describes a wave profile $g(\xi)$ that has 
a local maximum $g_0$ at $\xi=0$ modulo $L$ 
and nodal cusps at $\xi=\pm L/2$ modulo $L$ such that $g\to 0$ while $|g'|\to\infty$,
where 
\begin{equation}
L = 2\int_{0}^{g_0} \frac{g^{ (k-1)/2 }}{\sqrt{(g_0-g)A(g)}} \,dg >0
\end{equation}
is a size parameter.

\subsection{Solitary waves}\label{solitary}

To obtain a solitary wave solution $g(\xi)>0$, 
we need 
\begin{align}
& 0>C_1> C_1^*:\quad E>0
\label{eq:solitary-1}
\\
& E=V(g_0)=V(g_1)
\label{eq:solitary-2}
\\
& V'(g_0) =0, 
\quad
V''(g_0) <0, 
\label{eq:solitary-3}
\\
& V'(g_1) >0 
\label{eq:solitary-4}
\\
& V(g) \leq E \text{ for } 0<g_0\leq g\leq g_1 . 
\label{eq:solitary-5}
\end{align}

The third condition \eqref{eq:solitary-3} yields 
\begin{equation}
C_1 = (k+1)g_0^k - 2g_0 
\end{equation}
and $(k+1)kg_0^{k-1}<2$, 
which gives
\begin{equation}\label{eq:solitary_V}
V(g) = g^{k+1} -g^2 +(2g_0 -(k+1)g_0^k)g
\end{equation}
and
\begin{equation}\label{eq:solitary-g0}
0<g_0<g^*
\end{equation}
where $g^*$ is the point \eqref{eq:gstar} at which $V''(g)=0$. 
Then the first part of the second condition \eqref{eq:solitary-2} determines 
\begin{equation}\label{eq:solitary_E}
E = g_0^2(1- k g_0^{k-1})
\end{equation}
in terms of $g_0$. 
Thus, we have 
\begin{equation}
E-V(g) = (g-g_0)^2A(g)
\end{equation}
and 
\begin{equation}\label{eq:solitary_A}
A(g) = 1 - \frac{gS_{k}(g,g_0)- kg_0^k}{g-g_0}
= 1- kS_{k}(g,g_0) +g\partial_g S_{k}(g,g_0) . 
\end{equation}
Next, the second part of condition \eqref{eq:solitary-2} gives $A(g_1)=0$. 
This equation implicitly determines $g_1$ in terms of $g_0$, as follows. 
We consider $A(g)=0$. 
Since $V(g)$ has an inflection point at $g^*<1$, 
we know that $g>g^*$. 
For $g=1$, we note $S_{k}(1,g_0) <S_{k}(1,1)=k$ 
from properties \eqref{eq:S-property1} and \eqref{eq:S-limit}.
Hence, we have 
$A(1)-1=(kg_0^k-S_{k}(1,g_0))/(1-g_0) > k(g_0^k-1)/(1-g_0) = -kS_{k}(1,g_0) > k^2$,
which implies $A(1) >0$. 
But for $g\gg 1$, clearly we have $A(g)\sim -g^{k-1}<0$ 
from expression \eqref{eq:solitary_A}. 
This implies $A(g)$ changes sign in the interval $[1,\infty)$ 
and therefore has a root $g=g_1>1>g^*$. 
As a consequence, the fifth condition \eqref{eq:solitary-5} is then satisfied 
due to $A(g)\geq 0$ for $g_0\leq g\leq g_1$. 
Moreover, 
from expression \eqref{eq:solitary_V} combined with $g_0<g^*<1$, 
we see that  $E=V(g_0) = g_0^2(1-kg_0^{k-1})>0$. 
Hence, the first condition \eqref{eq:solitary-1} is satisfied. 
Finally, the fourth condition \eqref{eq:solitary-4} reduces to 
$S_{k}(g_1,g_0)>2/(k+1)$,
which holds due to $g_0<g^*$ as follows. 
From $A(g_1)=0$, 
we get $g_1S_{k}(g_1,g_0) = g_1-g_0(1-kg_0^{k-1})$,
and so 
$g_1(S_{k}(g_1,g_0) -2/(k+1)) = g_1(k-1)/(k+1)-g_0(1-kg_0^{k-1})$. 
Then, $g_0<g^*$ gives $1-kg_0^{k-1}>1-k{g^*}^{k-1} = (k-1)/(k+1)$, 
which yields $g_1(S_{k}(g_1,g_0) -2/(k+1)) > (g_1-g_0)(k-1)/(k+1)>0$,
whereby we have $S_{k}(g_1,g_0)>2/(k+1)$. 

Therefore, the preceding analysis shows that 
a solitary wave exists iff the constant $g_0$ satisfies the inequality \eqref{eq:solitary-g0}. 
Because $g_0$ is an asymptotic minimum of the wave profile $g(\xi)$ as $\xi\to \pm\infty$,
the solution integral \eqref{eq:ode_integral} must be expressed as 
\begin{equation}\label{eq:solitary_ode_integral}
\pm \int_{g}^{g_1} \frac{g^{ (k-1)/2 }}{(g-g_0)\sqrt{A(g)}} \,dg = \xi 
\end{equation}
with $g_0\leq g\leq g_1$  such that $A(g_1)=0$ where $g=g_1$
is a (global) maximum of the wave profile. 
The asymptotic tail has exponential decay $g\to g_0$,
since $g_0$ is a critical point of $V(g)$ but $V''(g_0)\neq 0$.

\subsection{Cusped solitary waves}

This type of solitary wave solution $g(\xi)\geq 0$ 
arises when 
\begin{align}
& 0>C_1\geq C_1^*:\quad E>0
\label{eq:cuspsolitary-1}
\\
& E=V(g_0)
\label{eq:cuspsolitary-2}
\\
& V'(g_0) =0,
\quad
V''(g_0)\leq 0
\label{eq:cuspsolitary-3}
\\
& V(g) \leq E \text{ for } 0\leq g\leq g_0 . 
\label{eq:cuspsolitary-4}
\end{align}

The two conditions \eqref{eq:cuspsolitary-2}--\eqref{eq:cuspsolitary-3} 
yield the energy expression \eqref{eq:solitary_E} 
and the potential energy expression \eqref{eq:solitary_V}, 
giving again the expression \eqref{eq:solitary_A} for $E-V(g)$. 
But the second part of the third condition \eqref{eq:cuspsolitary-3} 
now gives 
\begin{equation}\label{eq:cuspsolitary-g0}
0<g_0\leq g^*
\end{equation}
where $g^*$ is the point \eqref{eq:gstar} at which $V''(g)=0$. 
The fourth condition \eqref{eq:cuspsolitary-4} holds as follows. 
We note $E-V(g_0)=0$. 
Then we have 
$-V'(g)= (g_0-g)((k+1)S_k(g,g_0)-2) = (2g_0^{1-k}/k)(g_0-g)((g_0/g^*)^{k-1}S_k(g,g_0)-kg_0^{k-1}) \leq (2g_0^{1-k}/k)(g_0-g)(S_k(g,g_0)-kg_0^{k-1}) <0$
from property \eqref{eq:S-property1} combined with the limit \eqref{eq:S-limit}. 
This implies $V(g)$ is decreasing as $g$ decreases, 
and hence $E-V(g)\geq 0$ for $0\leq g\leq g_0$, 
since $V(g)$ has no critical point $g<g_0$. 

Therefore, a cusped solitary wave exists 
iff the inequality \eqref{eq:cuspsolitary-g0} holds. 
The solution integral \eqref{eq:ode_integral} is then given by 
\begin{equation}\label{eq:cuspedsolitary_ode_integral}
\pm \int_{0}^{g} \frac{g^{ (k-1)/2 }}{(g_0-g)\sqrt{A(g)}} \,dg = \xi 
\end{equation}
with $0\leq g\leq g_0$, 
where $A(g)$ is given by expression \eqref{eq:solitary_A}. 
This solution describes a wave profile $g(\xi)$ that has 
an asymptotic maximum $g_0$ as $\xi\to \pm\infty$, 
and a nodal cusp at $\xi=0$ such that $g\to 0$ while $|g'|\to\infty$. 

The behavior of the asymptotic tail $g\to g_0$ for large $\xi$
differs depending on whether $C_1=C_1^*$ or $C_1>C_1^*$. 
These two cases respectively correspond to $V''(g_0)=0$ and $V''(g_0)\neq 0$. 
Consequently, 
the general analysis using equation \eqref{eq:g-near-g0} shows that 
$g_0-g\to 0$ will exhibit exponential decay when $V''(g_0)\neq 0$, 
or power decay when $V''(g_0)=0$.

\subsection{Limiting case of periodic waves}\label{limiting}

A special limiting case of periodic waves arises when $g_0=0$.
This case is given by 
\begin{align}
& C_1\geq 0:\quad E=0
\label{eq:limiting-1}
\\
& V(0)=V(g_1)=0
\label{eq:limiting-2}
\\
& V'(g_1) \neq 0 
\label{eq:limiting-3}
\\
& V(g) \leq 0 \text{ for } 0\leq g\leq g_1
\label{eq:limiting-4}
\end{align}
where, now, the solution $g(\xi)$ will exhibit nodes at which $g=g_0=0$. 

The first and second conditions \eqref{eq:limiting-1}--\eqref{eq:limiting-2} 
together yield
\begin{equation}\label{eq:limiting_C1}
C_1 = g_1^{k} - g_1 \geq 0
\end{equation}
from which we get 
\begin{equation}\label{eq:limiting_V}
V(g) = g(g^k-g_1^k +g_1-g) . 
\end{equation}
This gives
\begin{equation}\label{eq:limiting_A}
E-V(g) = (g_1-g)gA(g),
\quad
A(g) = S_{k}(g,g_1)-1 . 
\end{equation}
Next, from the inequality \eqref{eq:limiting_C1}, 
we have 
\begin{equation}\label{eq:limiting-g1}
g_1 \geq 1 . 
\end{equation}
This implies that $V'(g_1) = g_1(kg_1^{k-1}-1)\geq k-1 > 0$ due to $k>1$,
and hence the third condition \eqref{eq:limiting-3} holds. 
The fourth condition \eqref{eq:limiting-4} now becomes $A(g)\geq 0$ 
for $0\leq g\leq g_1$. 
We first note, by direct evaluation,  
$A(0)= S_{k}(0,g_1)-1 = g_1^{k-1} -1>0$ 
due to inequality \eqref{eq:limiting-g1}. 
Next, we have $A'(g)= S_{k}'(g,g_1)>0$ by property \eqref{eq:S-property2'},
and hence we conclude $A(g)\geq A(0)$ for $0\leq g\leq g_1$. 
This shows that condition \eqref{eq:limiting-4} holds. 

Therefore, from this analysis, 
a periodic wave with $g_0=0$ exists iff the constant $g_1$ 
satisfies the inequality \eqref{eq:limiting-g1}. 
The solution integral \eqref{eq:ode_integral} is given by 
\begin{equation}\label{eq:limiting_ode_integral}
\pm \int_{0}^{g} \frac{g^{k/2-1}}{\sqrt{(g_1-g)A(g)}} \,dg = \xi 
\end{equation}
with $0\leq g\leq g_1$. 
The wave profile $g(\xi)$ for this solution \eqref{eq:limiting_ode_integral}
will be periodic. 
It will have a local maximum of $g_1$ at $\xi=\pm L/2$ modulo $L$,
where its wavelength is given by 
\begin{equation}
L = 2\int_{0}^{g_1} \frac{g^{k/2-1}}{\sqrt{(g_1-g)A(g)}} \,dg >0 . 
\end{equation}
At $\xi=0$ modulo $L$, 
the wave will have either a cusp, a corner, or a local minimum, 
depending on $g_1$, 
as indicated in Table~\ref{table:types}
which follows from inequality \eqref{eq:limiting_C1}.

\subsection{Energy and momentum}

Each solution $g(\xi)$ of the first-order ODE \eqref{eq:ode_energy_form} 
gives a corresponding solution of the strain ODE \eqref{eq:g-ode} 
after we revert to the physical (unscaled) variables $\g(\tilde\xi)$. 
From the scaling transformation \eqref{eq:scal-g-xi}--\eqref{eq:scaling}, 
we get 
\begin{equation}\label{eq:phys-g-xi}
\g(\tilde\xi)  = \left(\tfrac{1}{2} (\V/c)^2 k(k+1)\right)^{1/(k-1)} 
g\Big(\tfrac{\sqrt{6}}{\sqrt{k(k+1)}R}\tilde\xi\Big)
\end{equation}
which represents a physical travelling wave in terms of the strain variable 
\begin{equation}\label{eq:v-soln}
v= \g(x-\V t) 
= \left(\tfrac{1}{2} (\V/c)^2 k(k+1)\right)^{1/(k-1)} 
g\Big(\tfrac{\sqrt{6}}{\sqrt{k(k+1)}R}(x-\V t)\Big)
\end{equation}
where $\V=\const$ is the wave speed. 
Note that these waves can be bi-directional, 
since there is no constraint on the sign of $\V$, 
other than $\V\neq0$. 

All physical travelling waves carry energy and momentum, 
which are defined by the respective conservation laws \eqref{eq:TX3} and \eqref{eq:TX4}
where $\u_t=-\V\g$ and $\u_x=\g$. 
The total energy and total momentum are given by 
the spatial integrals \eqref{eq:C3} and \eqref{eq:C4} on a domain 
$\Omega = (-\infty,\infty)$ in the case of solitary waves,  
or $\Omega = (-\L/2,\L/2)$ in the case of periodic waves with wavelength $\L>0$. 
In terms of $\g(\tilde\xi)$, these integrals take the form 
\begin{align}
& \P= C_3[\g]=\int_{\Omega} \V\g^2\,d\tilde\xi,
\label{eq:v_P}
\\
& \E=C_4[\g]=\int_{\Omega}
\big( \tfrac{1}{2} \V^2 \g^2 - c^2 ( \tfrac{1}{6}R^2 \g^{k-1} (\g')^2 - \tfrac{1}{k(k+1)} \g^{k+1} ) \big) \,d\tilde\xi . 
\label{eq:v_E}
\end{align}
For periodic waves, both integrals will be finite, since the domain is a finite interval. 

But for solitary waves, the domain is infinite and $\g(\tilde\xi)$ has a non-zero asymptotic value $\g_0>0$ as $\tilde\xi\to \pm\infty$, 
and hence both integrals will diverge. 
However, we can regularize the energy density and momentum density 
to remove the divergent contribution in two different ways. 

Firstly, 
we can consider a constant solution given by $\g=\g_0$,
which satisfies the strain ODE \eqref{eq:g-ode},
where $\g_0$ is the asymptotic value of a solitary wave solution. 
The respective momentum and energy densities \eqref{eq:TX3} and \eqref{eq:TX4}
evaluated for the corresponding constant strain solution $v=\g_0$ are given by 
\begin{align}
& T_3[\g_0]=\V\g_0^2 ,
\label{eq:T3-g0}
\\
& T_4[\g_0]= \tfrac{1}{2} \V^2 \g_0^2 + \tfrac{1}{k(k+1)} c^2 \g_0^{k+1} . 
\label{eq:T4-g0}
\end{align}
If we now subtract these densities from the respective densities 
in the integrals \eqref{eq:v_P} and \eqref{eq:v_E},
then we obtain regularized momentum and energy integrals 
\begin{align}
& \tilde\P= \int_{\Omega}\V (\g^2-\g_0^2) \,d\tilde\xi , 
\label{eq:v_P_reg}
\\
& \tilde\E =\int_{\Omega} \big( \tfrac{1}{2}\V^2(\g^2-\g_0^2) - c^2 ( \tfrac{1}{6}R^2 \g^{k-1} (\g')^2 - \tfrac{1}{k(k+1)} (\g^{k+1} - \g_0^{k+1} ) ) \big) \,d\tilde\xi , 
\label{eq:v_E_reg}
\end{align}
both of which will be finite.  

Secondly, 
we can instead consider a linear combination of the energy and momentum densities 
such that the resulting density goes to zero as $\tilde\xi\to \pm\infty$. 
This is achieved simply by taking the density to be $T_4[\g] -(T_4[\g_0]/T_3[\g_0])T_3[\g]$.
From expressions \eqref{eq:T3-g0} and \eqref{eq:T4-g0}, 
we then have the energy-momentum integral 
\begin{equation}\label{eq:v_EP}
\begin{aligned}
\hat\E =\int_{\Omega} c^2\big(  \tfrac{1}{k(k+1)} \g^2(\g^{k-1} - \g_0^{k-1})  - \tfrac{1}{6}R^2 \g^{k-1} (\g')^2 \big) \,d\tilde\xi
\end{aligned}
\end{equation}
which will be finite.  

A direct evaluation of any of these integrals 
\eqref{eq:v_P}, \eqref{eq:v_E}, \eqref{eq:v_P_reg}, \eqref{eq:v_E_reg}, \eqref{eq:v_EP}
would require that 
we first substitute the explicit form of the travelling wave solution $\g(\tilde\xi)$
and then integrate with respect to $\tilde\xi$ over the domain $\Omega$. 
We can avoid these obstacles by 
expressing each integral in an equivalent form \eqref{eq:conservedquantity} 
which involves only the ODE for the travelling wave. 
In particular, 
we first use the scaling transformation \eqref{eq:phys-g-xi}
to express $\g(\tilde\xi)$ in terms of the dimensionless expression $g(\xi)$, 
and we next use the dimensionless first-order ODE \eqref{eq:ode_energy_form} 
to change the integration variable from $\xi$ to $g$. 
This formulation relies on the global features of $g(\xi)$ 
that all travelling waves are symmetric functions of $\xi$, 
and that non-nodal periodic waves and solitary waves 
have a single maximum amplitude $g_{\max}$ at the center of the domain $\Omega$ 
and a minimum amplitude $g_{\min}>0$ at the domain endpoints $\partial\Omega$,
whereas cusped and peaked waves as well as nodal periodic waves 
have a single minimum amplitude $g_{\min}=0$ at the center of the domain $\Omega$ 
and a maximum amplitude $g_{\max}>0$ at the domain endpoints $\partial\Omega$. 

Thus, the momentum and energy integrals \eqref{eq:v_P}--\eqref{eq:v_E}
for all periodic waves, including the case of cusped and peaked periodic waves, 
can be expressed as 
\begin{align}
& \P = \tfrac{1}{\sqrt{3}} \sigma^{(k+3)/(k-1)}|\V/c|^{4/(k-1)} \V R 
\int_{g_{\min}}^{g_{\max}} \frac{g^{(k+3)/2}}{\sqrt{E-V(g)}}\,d g , 
\label{eq:periodic-P}
\\
& \E= \tfrac{1}{2\sqrt{3}} \sigma^{(k+3)/(k-1)}|\V/c|^{4/(k-1)} \V^2 R 
\int_{g_{\min}}^{g_{\max}} g^{(k-1)/2}\bigg( \frac{g^2 (1+g^{k-1})}{\sqrt{E-V(g)}} -\sqrt{E-V(g)} \bigg)\,d g , 
\label{eq:periodic-E}
\end{align}
with 
\begin{equation}
\sigma = \sqrt{\tfrac{1}{2}k(k+1)} .
\end{equation} 
We remark that the physical wavelength $\L$ of a periodic wave 
is related to the dimensionless wavelength $L$ of $g(\xi)$ 
by 
\begin{equation}
\L = \sqrt{\tfrac{k(k+1)}{6}} L R . 
\end{equation}

Similarly, 
for all solitary waves,
including the case of cusped solitary waves, 
the regularized momentum and energy integrals \eqref{eq:v_P_reg}--\eqref{eq:v_E_reg}
along with the energy-momentum integral \eqref{eq:v_EP}
can be expressed as 
\begin{align}
& \tilde\P = \tfrac{1}{\sqrt{3}} \sigma^{(k+3)/(k-1)}|\V/c|^{4/(k-1)} \V R 
\int_{g_{\min}}^{g_{\max}} \frac{(g^2-g_0^2)g^{(k-1)/2}}{\sqrt{E-V(g)}}\,d g, 
\label{eq:solitary-P}
\\
& \tilde\E= \tfrac{1}{2\sqrt{3}} \sigma^{(k+3)/(k-1)}|\V/c|^{4/(k-1)} \V^2 R 
\int_{g_{\min}}^{g_{\max}} g^{(k-1)/2}\bigg( \frac{g^2 -g_0^2 +g^{k+1}-g_0^{k+1}}{\sqrt{E-V(g)}} -\sqrt{E-V(g)} \bigg)\,d g, 
\label{eq:solitary-E}
\end{align}
and  
\begin{equation}
\hat\E= \tfrac{1}{2\sqrt{3}} \sigma^{(k+3)/(k-1)}|\V/c|^{4/(k-1)} \V^2 R 
\int_{g_{\min}}^{g_{\max}} g^{(k-1)/2}\bigg( \frac{g^2(g^{k-1}-g_0^{k-1})}{\sqrt{E-V(g)}} -\sqrt{E-V(g)} \bigg)\,d g.
\label{eq:solitary-EP}
\end{equation}

All of these integrals \eqref{eq:periodic-P}--\eqref{eq:periodic-E}
and \eqref{eq:solitary-P}--\eqref{eq:solitary-EP} 
are finite and represent physical conserved quantities for travelling waves.

\section{Explicit solutions}\label{sec:examples}

The different solution integrals \eqref{eq:ode_integral} for 
periodic waves \eqref{eq:periodic_ode_integral}, 
solitary waves \eqref{eq:solitary_ode_integral},
cusped periodic waves \eqref{eq:cuspedperiodic_ode_integral}, 
cusped solitary waves \eqref{eq:cuspedsolitary_ode_integral}, 
and nodal waves \eqref{eq:limiting_ode_integral}
can be evaluated in an explicit analytical form
in various cases which will now be systematically determined. 

For a given solution integral, our aim is to find a suitable change of variable 
\begin{equation}\label{eq:h}
g=h^q, 
\quad 
q>0
\end{equation}
that will bring the integral to one of the forms 
\begin{equation}\label{eq:integral-form}
\int_{g^{1/q}}^{g_{\max}^{1/q}} \frac{F(h)}{\sqrt{P(h)}} \,dh 
\quad\text{ or }\quad
\int_{g_{\min}^{1/q}}^{g^{1/q}} \frac{F(h)}{\sqrt{P(h)}} \,dh
\end{equation}
with $g_{\min}\leq g\leq g_{\max}$, 
where $F(h)$ is a rational function 
and $P(h)$ is a polynomial of lowest possible degree $d>0$. 
If $d$ is at most two, then the integral can be evaluated explicitly 
in terms of elementary functions; 
if $d$ is three or four, then the integral can be evaluated explicitly
in terms of elliptic functions. 

We will now classify all of the cases for which this method produces 
an explicit evaluation of each solution integral 
\eqref{eq:periodic_ode_integral}, \eqref{eq:solitary_ode_integral},
\eqref{eq:cuspedperiodic_ode_integral}, \eqref{eq:cuspedsolitary_ode_integral}, 
\eqref{eq:limiting_ode_integral}. 
This classification depends sensitively on the nonlinearity exponent $k>1$.

\subsection{Classification of explicit solution integrals}\label{sec:classification}

Applying the change of variable \eqref{eq:h} to the general solution integral \eqref{eq:ode_integral}, 
we obtain 
\begin{equation}\label{eq:soln-integral}
\int_{g_0}^{g} \frac{g^{(k-1)/2}}{\sqrt{E-V(g)}} \,dg 
= \int_{h_0}^{h} \frac{q h^{q(k+1)/2 -1}}{\sqrt{E-V(h^q)}} \,dh  . 
\end{equation}
We can bring this integral to the general form \eqref{eq:integral-form}
by multiplying the numerator and denominator by $h^r$ 
where $q$ and $r$ are chosen to make 
$h^{q(k+1)/2 +r-1}$ be a monomial and $h^{2r}(E-V(h^q))$ be a polynomial of lowest possible degree. 
We then have 
$F(h) = h^{q(k+1)/2 +r-1}$ 
and 
$P(h) = h^{2r}(E-V(h^q))$
in all cases when $E-V(h^q)$ has no roots with a degeneracy greater than one. 
This encompasses all of the different types of periodic waves
(\cf/ Table~\ref{table:types}). 
In the case when $E-V(h^q)$ has a root $h=h_0$ with a degeneracy of two or three, 
we can factorize $E-V(h^q)= (h-h_0)^2 B(h)$ to get 
$F(h) = h^{q(k+1)/2 +r-1}/|h-h_0|$ 
and 
$P(h) = h^{2r}B(h)$. 
This encompasses all of the types of solitary waves 
(\cf/ Table~\ref{table:types}). 
We note that if $E\neq 0$ then the lowest power term in $h^{2r}(E-V(h^q))$ is $Eh^{2r}$
and so in this case we can assume $2r=0$ or $2r=1$ 
(since otherwise $h^{2\lfloor r\rfloor}$ can be factored out to get 
$P(h)=E-V(h^q)$ when $2r$ is even or $P(h)=Eh-hV(h^q)$ when $2r$ is odd, 
which are polynomials of lower degree). 

We will now proceed to classify all possibilities for $q$ and $r$ 
such that $P(h)$ has degree at most two or at most four. 
If more than one possibility works for a given value of $k>1$, 
then we will choose the one that yields the lowest degree for $P(h)$. 
This will yield a classification of all distinct cases 
for which the solution integral \eqref{eq:soln-integral} can be evaluated explicitly 
in terms of elementary functions or elliptic functions. 

Note from equation \eqref{eq:gstar} we have 
\begin{equation}\label{eq:hstart}
h^* = (g^*)^{1/q} = \big( \tfrac{2}{k(k+1)}\big)^{1/(q(k-1))} . 
\end{equation}

\subsubsection{Solitary waves}\label{sec:solitary}
We begin with solitary waves \eqref{eq:solitary_ode_integral}. 
The most general form \eqref{eq:integral-form} 
for the solution integral \eqref{eq:soln-integral} in this case
is given by 
\begin{gather}
F(h)=  \frac{q h^p}{h-h_0}
\\
(h-h_0)^2P(h) =  E h^{2r} +C_1 h^{2r+q} + h^{2r+2q} - h^{2p+2}
\end{gather}
with
\begin{equation}
p= \tfrac{1}{2}q(k+1) + r-1, 
\quad
E = h_0^{2q}(1-k h_0^{q(k-1)}) , 
\quad
C_1 =  -h_0^{q}(2-(k+1) h_0^{q(k-1)}) . 
\end{equation}
Note we have $E>0$ and $C_1<0$ due to inequality \eqref{eq:solitary-g0}. 
Consequently, we can assume $r=0$ or $r=1/2$. 
Then, 
$F(h)$ will be a rational function iff 
$p$ is an integer, 
and $P(h)$ will be a polynomial of degree $d\leq 4$ iff 
$Eh^{2r} +C_1 h^{2r+q} + h^{2r+2q} - h^{2p+2}$ is a polynomial of degree at most six,
where we have $2r<2r+q<2r+2q<2p+2$ due to $k>1$ and $q>0$. 
This determines the following classification:
\begin{subequations}
\begin{align}
2p+2 & = 4,\quad 4,\quad 6,\quad 6,\quad 6,\quad 6
\\
2r +2q & = 2,\quad 3,\quad 2,\quad 4,\quad 3,\quad 5
\\
2r+q & = 1,\quad 2,\quad 1,\quad 2,\quad 2,\quad 3\quad
\\
2r & = 0,\quad 1,\quad 0,\quad 0,\quad 1,\quad 1
\end{align}
\end{subequations}
where the corresponding values of $d=2p$ and $k=2(p+1-r)/q -1$ are given by
\begin{align}
d & = 2,\quad 2,\quad 4,\quad 4,\quad 4,\quad 4
\\
k & = 3,\quad 2,\quad 5,\quad 2,\quad 4,\quad \tfrac{3}{2}
\end{align}
The only cases that have $d\leq 2$ are when $k=2$ and $k=3$,
where $q=1$ in both cases. 

For $k=3$, 
we have $F(h)=h/(h-h_0)$ and $P(h)=(h_1-h)(h_1+2h_0+h)$, 
where $h_1 = \sqrt{1-2h_0^2}-h_0$ with $h_0<h^*=1/\sqrt{6}$. 
The solution integral \eqref{eq:integral-form} in this case yields
\begin{equation}
\int_{h}^{h_1}\frac{F(h)}{\sqrt{P(h)}}\,dh 
= \tfrac{1}{2}\pi -\arctan\bigg( \frac{h + h_0}{\sqrt{P(h)}} \bigg) 
+  \frac{h_0}{\sqrt{P(h_0)}} \arctanh\Big(Q(h)\Big) 
\end{equation}
where
\begin{equation}\label{eq:Q}
Q(h)=\frac{2\sqrt{P(h)P(h_0)}}{P(h)+ P(h_0)+(h-h_0)^2} . 
\end{equation}

For $k=2$, 
we have $F(h)=h/(h-h_0)$ and $P(h)=(h_1-h)h$, 
where $h_1 = 1-2h_0$ with $h_0<h^*=1/3$. 
The solution integral \eqref{eq:integral-form} in this case yields
\begin{equation}
\int_{h}^{h_1}\frac{F(h)}{\sqrt{P(h)}}\,dh 
= \tfrac{1}{2}\pi -\arctan\bigg( \frac{h + h_0-\tfrac{1}{2}}{\sqrt{P(h)}} \bigg) 
+ \frac{h_0}{\sqrt{P(h_0)}} \arctanh\Big( Q(h) \Big)  . 
\end{equation}

Hence, 
using $h=g$ since $q=1$, 
and taking into account the necessary and sufficient condition \eqref{eq:solitary-g0} 
on $g_0$, 
we obtain the following result. 

\begin{prop}\label{prop:solitary}
(i) The solitary wave solution integral \eqref{eq:solitary_ode_integral} 
has an explicit evaluation in terms of elementary functions when (and only when) 
$k=2,3$. 
These solutions are given by algebraic equations
\begin{equation}
\begin{aligned}
k=2:\quad& 
\tfrac{1}{2}\pi -\arctan\bigg( \frac{g + g_0-\tfrac{1}{2}}{\sqrt{P(g)}} \bigg) 
+ \frac{g_0}{\sqrt{P(g_0)}} \arctanh\Big(Q(g)\Big) 
=\pm\xi,
\\& 
P(g) = (g_1-g)g, 
\quad
g_1 = 1-2g_0,
\quad
g_0<g^*=1/3
\end{aligned}
\label{eq:solitarysol1}
\end{equation}
and
\begin{equation}
\begin{aligned}
k=3:\quad&
\tfrac{1}{2}\pi -\arctan\bigg( \frac{g + g_0}{\sqrt{P(g)}} \bigg) 
+  \frac{g_0}{\sqrt{P(g_0)}} \arctanh\Big(Q(g)\Big) 
=\pm \xi ,
\\& 
P(g) = (g_1-g)(g+g_1+2g_0),
\quad
g_1 = \sqrt{1-2g_0^2}-g_0,
\quad
g_0<g^*=1/\sqrt{6}
\end{aligned}
\label{eq:solitarysol2}
\end{equation}
where $Q(g)$ is given by expression \eqref{eq:Q}. \\
(ii) The solitary wave solution integral \eqref{eq:solitary_ode_integral} 
has an explicit evaluation in terms of elliptic functions when (and only when) 
$k=\tfrac{3}{2},4,5$. 
\end{prop}

\begin{figure}[!h]
\centering
\includegraphics[width=\textwidth]{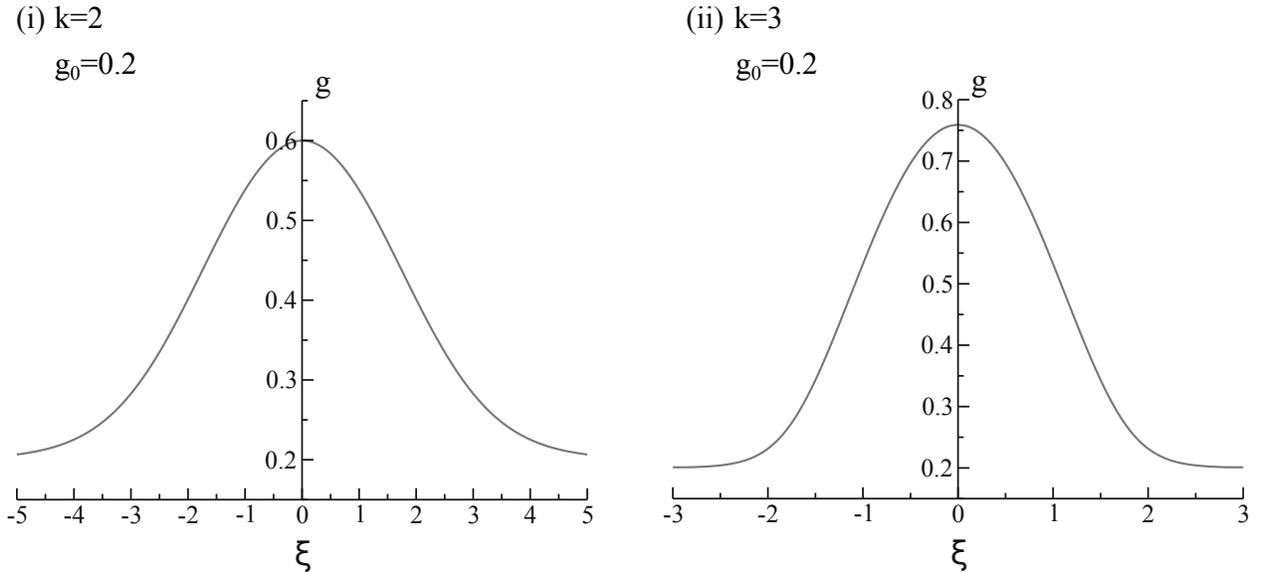}
\caption{Exact solitary wave solutions.}
\label{fig:Solitary-solutions}
\end{figure}

\subsubsection{Cusped solitary waves}\label{sec:cuspedsolitary}
The classification for cusped solitary waves \eqref{eq:cuspedsolitary_ode_integral} 
is the same, apart from a change in the sign of the denominator of $F(h)$
and a change in the integration domain. 
This leads to the following result.

\begin{prop}\label{prop:cuspedsolitary}
(i) The cusped solitary wave solution integral \eqref{eq:cuspedsolitary_ode_integral} 
has an explicit evaluation in terms of elementary functions when (and only when) 
$k=2,3$. 
These solutions are given by algebraic equations
\begin{equation}
\begin{aligned}
k=2:\quad& 
\arctan\bigg( \frac{\frac{1}{2}-(g + g_0)}{\sqrt{P(g)}} \bigg) -\tfrac{1}{2}\pi 
+ \frac{g_0}{\sqrt{P(g_0)}} \ln\bigg(\frac{R(g)}{(g_0-g)(1-2g_0)}\bigg) 
=\pm \xi,
\\& 
P(g) = (g_1-g)g, 
\quad
g_1 = 1-2g_0,
\quad
g_0<g^*=1/3
\end{aligned}
\label{eq:cuspedsolsol1}
\end{equation}
and
\begin{equation}
\begin{aligned}
k=3:\quad& 
\arctan\bigg( \frac{g_0\sqrt{P(g)}-(g +  g_0)\sqrt{P(0)}}{g_0(g+g_0) + \sqrt{P(g)P(0)}} \bigg) 
+  \frac{g_0}{\sqrt{P(g_0)}} \ln \left ( \frac{g_0R(g)}{(g_0-g)R(0)} \right) 
=\pm\xi,
\\& 
P(g) = (g_1-g)(g+g_1+2g_0),
\quad
g_1 = \sqrt{1-2g_0^2}-g_0,
\quad
g_0<g^*=1/\sqrt{6}
\end{aligned}
\label{eq:cuspedsolsol2}
\end{equation}
where
\begin{equation}
\label{eq:R}
R(g) = (\sqrt{P(g)}+\sqrt{P(g_0)})^2 + (g-g_0)^2. 
\end{equation}
(ii) The cusped solitary wave solution integral \eqref{eq:solitary_ode_integral} 
has an explicit evaluation in terms of elliptic functions when (and only when) 
$k=\tfrac{3}{2},4,5$. 
\end{prop}

\begin{figure}[!h]
\centering
\includegraphics[width=\textwidth]{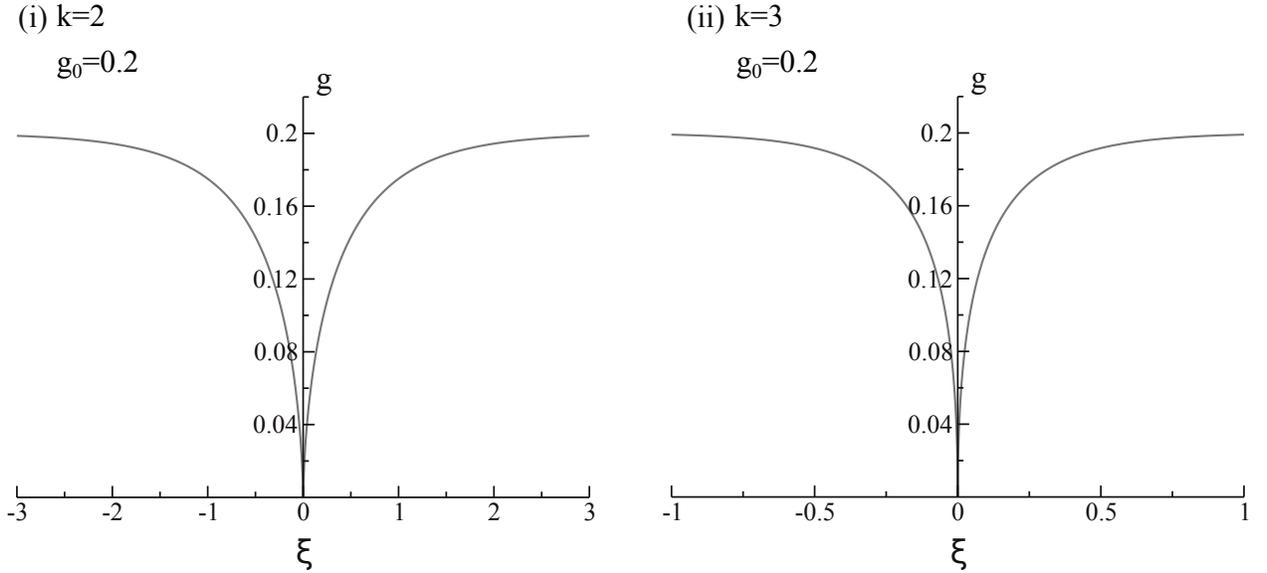}
\caption{Exact cusped solitary wave solutions.}
\label{fig:Cusped-Solitary-solutions}
\end{figure}

\subsubsection{Periodic waves}\label{sec:periodic}
We next look at periodic waves \eqref{eq:periodic_ode_integral}. 
In this case the most general form \eqref{eq:integral-form} 
for the solution integral \eqref{eq:soln-integral} 
is given by 
\begin{gather}
F(h)=  q h^p
\\
P(h) = E h^{2r} +C_1 h^{2r+q} + h^{2r+2q} - h^{2p+2}
\end{gather}
with
\begin{equation}
p= \tfrac{1}{2}q(k+1) +r -1, 
\quad
E = h_0^q h_1^q (1-S_{k}(h_1^q,h_0^q)) , 
\quad
C_1 = S_{k+1}(h_1^q,h_0^q)) -h_1^q - h_0^q . 
\end{equation}
Here we can have $E=0$ or $C_1=0$ but not both $E=C_1=0$ 
(since this leads to $g_0=1$ or $g_1=1$, both of which violate inequality \eqref{eq:periodic-g0g1}). 
Thus there will be three subcases to consider:
$E=0$, $C_1\neq 0$; 
$E\neq 0$, $C_1=0$; 
$E\neq 0$, $C_1\neq 0$. 

When $E\neq 0$ and $C_1\neq 0$, 
$F(h)$ will be a rational function iff 
$p$ is an integer, 
and $P(h)$ will be a polynomial of degree $d\leq 4$ iff 
$2r<2r+q<2r+2q<2p+2 \leq 4$ are integers, 
where we can assume $r=0$ or $r=1/2$. 
This determines the following classification:
\begin{subequations}
\begin{align}
2p+2 & = 4,\quad 4
\\
2r +2q & = 2,\quad 3
\\
2r+q & = 1,\quad 2
\\
2r & = 0,\quad 1
\end{align}
\end{subequations}
where the corresponding values of $d=2p+2$ and $k=2(p+1-r)/q -1$ are given by
\begin{align}
d & = 4,\quad 4
\\
k & = 3,\quad 2
\end{align}
There are no cases that have $d\leq 2$. 

When $E=0$ and $C_1\neq 0$, 
$F(h)$ will be a rational function iff 
$p$ is an integer, 
and $P(h)$ will be a polynomial of degree $d\leq 4$ iff 
$2r+q<2r+2q<2p+2\leq 4$ are integers. 
Note we can consequently assume $2r+q=0$ or $2r+q=1$. 
This determines the following classification:
\begin{subequations}
\begin{align}
2p+2 & = 2,\quad 4,\quad 4,\quad 4,\quad 4,\quad 4
\\
2r +2q & = 1,\quad 1,\quad 2,\quad 3,\quad 2,\quad 3
\\
2r+q & = 0,\quad 0,\quad 0,\quad 0,\quad 1,\quad 1
\end{align}
\end{subequations}
where the corresponding values of $d=2p+2$ and $k=2(p+1-r)/q-1$ are given by
\begin{align}
d & = 2,\quad 4,\quad 4,\quad 4,\quad 4,\quad 4 
\\
k & = 2,\quad 4,\quad 2,\quad \tfrac{4}{3},\quad 3,\quad \tfrac{3}{2}
\end{align}
The only case that has $d\leq 2$ is when $k=2$,
where $q=1$. 
Then we have $F(h)=1$ and $P(h)=(h_1-h)(h-h_0)$
where $h_1=1-h_0$. 
The solution integral \eqref{eq:integral-form} in this case yields
\begin{equation}
\int_{h_0}^{h}\frac{F(h)}{\sqrt{P(h)}}\,dh 
= \arctan\bigg( \frac{\frac{1}{2}-h}{\sqrt{P(h)}} \bigg) + \tfrac{1}{2} \pi . 
\end{equation}

When $E\neq 0$ and $C_1=0$, 
$F(h)$ will be a rational function iff 
$p$ is an integer, 
and $P(h)$ will be a polynomial of degree $d\leq 4$ iff 
$2r<2r+2q<2p+2\leq 4$ are integers. 
Note we can assume $r=0$ or $r=1/2$. 
This determines the following classification:
\begin{subequations}
\begin{align}
2p+2 & = 2,\quad 4,\quad 4,\quad 4,\quad 4,\quad 4
\\
2r +2q & = 1,\quad 1,\quad 2,\quad 3,\quad 2,\quad 3
\\
2r & = 0,\quad 0,\quad 0,\quad 0,\quad 1,\quad 1
\end{align}
\end{subequations}
where the corresponding values of $d=2p+2$ and $k=2(p+1-r)/q-1$ are given by
\begin{align}
d & = 2,\quad 4,\quad 4,\quad 4,\quad 4,\quad 4 
\\
k & = 3,\quad 7,\quad 3,\quad \tfrac{5}{3},\quad 5,\quad 2
\end{align}
The only case that has $d\leq 2$ is when $k=3$,
where $q=1/2$.
Then we have $F(h)=1/2$ and $P(h)=(h_1-h)(h-h_0)$
where $h_0=1-h_1$. 
In this case, the solution integral \eqref{eq:integral-form} yields
\begin{equation}
\int_{h_0}^{h}\frac{F(h)}{\sqrt{P(h)}}\,dh 
= \tfrac{1}{2}\arctan\bigg( \frac{\frac{1}{2}-h}{\sqrt{P(h)}} \bigg) + \tfrac{1}{4} \pi .
\end{equation}

Hence, 
using $h=g$ when $k=2$ since $q=1$, 
and $h=g^2$ when $k=3$ since $q=1/2$, 
we obtain the following result
after taking into account the necessary and sufficient conditions \eqref{eq:periodic-Sk}, \eqref{eq:periodic-g0g1}, \eqref{eq:periodic-g1gstar}
on $g_1$ and $g_0$. 

\begin{prop}\label{prop:periodic}
(i) The periodic wave solution integral \eqref{eq:periodic_ode_integral} 
has an explicit evaluation in terms of elementary functions when (and only when) 
$k=2$ with $E=0$ and $C_1\neq 0$,  
and $k=3$ with $C_1=0$ and $E\neq 0$. 
These solutions are given by the explicit expressions
\begin{equation}
\begin{aligned}
k=2:\quad& 
g= \tfrac{1}{2}\Big( 1 +(2g_1-1)\cos(\xi)\Big), 
%\arctan\Big( \frac{\frac{1}{2}-g}{\sqrt{P(g)}} \Big) + \tfrac{1}{2} \pi =\pm (\xi-\pi),
% P(g) = (g_1-g)(g-g_0), \quad g_0=1-g_1, 
\quad
1>g_1 >1/2
\\& L=2\pi
\end{aligned}
\label{eq:periodicsol1}
\end{equation}
and
\begin{equation}
\begin{aligned}
k=3:\quad& 
g= \sqrt{\tfrac{1}{2}\Big( 1 +(2g_1^2-1)\cos(2\xi)\Big)}, 
% \arctan\Big( \frac{\frac{1}{2}-g^2}{\sqrt{P(g^2)}} \Big) + \tfrac{1}{2} \pi =\pm 2(\xi-\pi/2),
% P(g^2) = (g_1^2-g^2)(g^2-g_0^2), \quad g_0=\sqrt{1-g_1^2}, 
\quad
1>g_1 >1/\sqrt{2}
\\& L=\pi
\end{aligned}
\label{eq:periodicsol2}
\end{equation}
where $L$ denotes the wavelength. \\
(ii) The periodic wave solution integral \eqref{eq:periodic_ode_integral} 
has an explicit evaluation in terms of elliptic functions when (and only when) 
$k=2,3$ with $E\neq 0$ and $C_1\neq 0$; 
$k=\tfrac{5}{3},2,5,7$ with $E\neq 0$ and $C_1=0$; 
$k=\tfrac{4}{3},\tfrac{3}{2},3,4$ with $E= 0$ and $C_1\neq 0$. 
\end{prop}

\begin{figure}[!h]
\centering
\includegraphics[width=\textwidth]{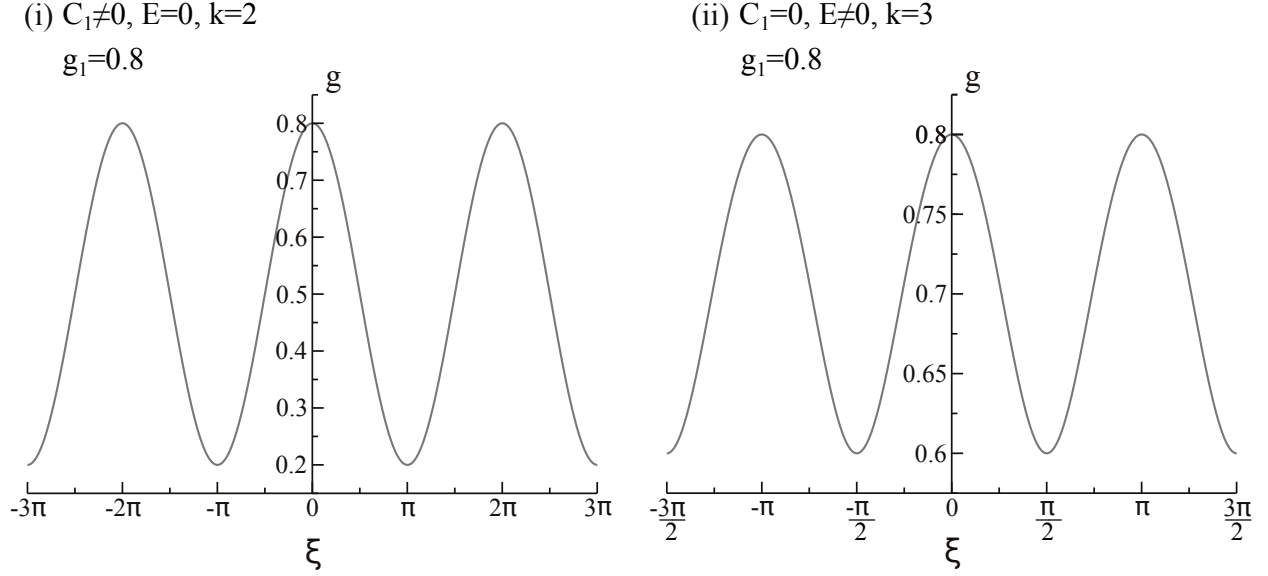}
\caption{Exact nonlinear periodic wave solutions.}
\label{fig:Periodic-solutions}
\end{figure}

\subsubsection{Cusped periodic waves}\label{sec:cuspedperiodic}
We continue with cusped periodic waves \eqref{eq:cuspedperiodic_ode_integral}. 
The most general form \eqref{eq:integral-form} 
for the solution integral \eqref{eq:soln-integral} 
is given by 
\begin{gather}
F(h)=  q h^p
\\
P(h) = E h^{2r} -C_1 h^{2r+q} + h^{2r+2q} - h^{2p+2}
\end{gather}
with
\begin{equation}
p= \tfrac{1}{2}(2r + q(k+1))-1, 
\quad
C_1 = Eh_0^{-q} + h_0^q - h_0^{kq},
\quad
E>0 . 
\end{equation}
Here we can have $C_1=0$, 
which gives two subcases to consider:
$C_1=0$; $C_1\neq 0$. 

The classification of these two cases is the same as the corresponding two cases 
$C_1=0$ and $C_1\neq 0$ for periodic waves with $E\neq 0$. 
This leads to the following result. 

\begin{prop}\label{prop:cuspedperiodic}
(i) The cusped periodic wave solution integral \eqref{eq:cuspedperiodic_ode_integral} 
has an explicit evaluation in terms of elementary functions when (and only when) 
$k=3$ with $C_1=0$. 
In this case the solution is given by the algebraic equation
\begin{equation}
\begin{aligned}
k=3:\quad& 
g= \sqrt{\tfrac{1}{2}\Big( 1 +(2g_0^2-1)\cos(2\xi-2L\lfloor \tfrac{1}{2} + \xi/L\rfloor)\Big)}, 
%\arctan\bigg( \frac{\frac{1}{2}-g^2}{\sqrt{P(g^2)}} \bigg) +\tfrac{1}{2}\pi =\pm 2\xi,
% P(g^2) = (g^2-1+g_0^2)(g_0^2-g^2), \quad 1>g_1 >1/\sqrt{2}
\quad
g_0>1, 
\\& L= \tfrac{1}{2}\pi +\arctan\bigg( \frac{1}{2g_0\sqrt{g_0^2-1}}\bigg)
\end{aligned}
\label{eq:cuspedpersol1}
\end{equation}
where $L$ denotes the wavelength, 
and $\lfloor x \rfloor$ denotes the floor function. \\
(ii) The cusped periodic wave solution integral \eqref{eq:periodic_ode_integral} 
has an explicit evaluation in terms of elliptic functions when (and only when) 
$k=2,3$ with $C_1\neq 0$, 
and $k=\tfrac{5}{3},2,5,7$ with $C_1=0$. 
\end{prop}

\begin{figure}[!h]
\centering
\includegraphics[width=\textwidth]{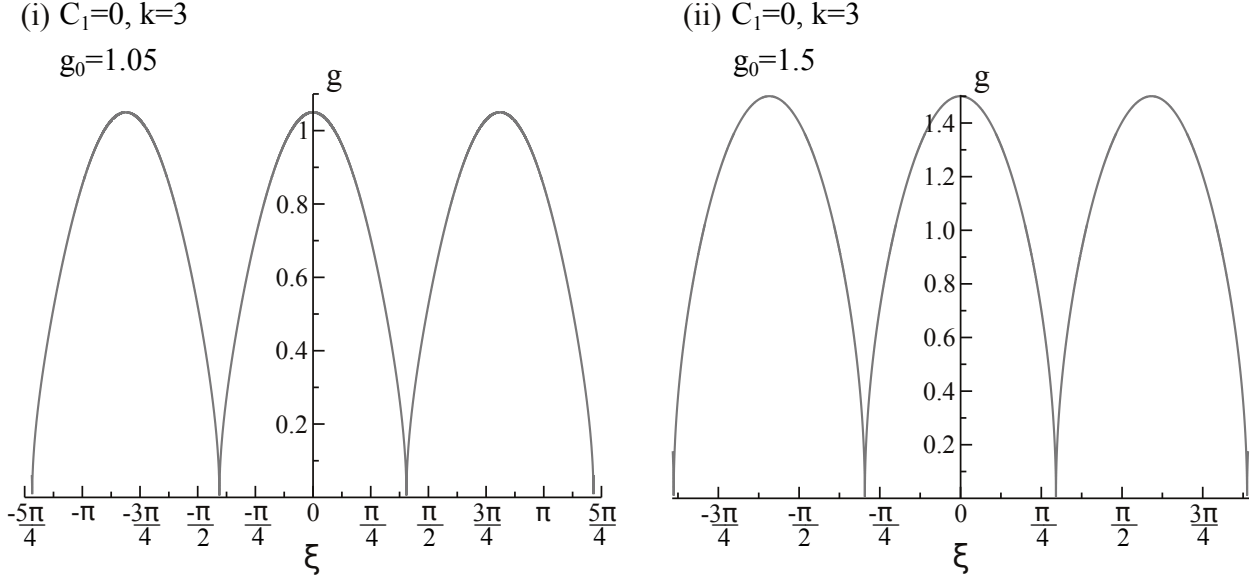}
\caption{Exact cusped periodic wave solutions.}
\label{fig:Cusped-Periodic-solutions}
\end{figure}

\subsubsection{Nodal waves}\label{sec:nodal}
Finally, we look at nodal waves \eqref{eq:limiting_ode_integral}. 
The most general form \eqref{eq:integral-form} 
for the solution integral \eqref{eq:soln-integral} 
splits into two different subcases:
$C_1 =0$ and $C_1\neq 0$. 

When $C_1\neq 0$, 
we have 
\begin{gather}
F(h)=  q h^p
\\
P(h) = C_1 h^{2r} + h^{2r+q} - h^{2p+2}
\end{gather}
with
\begin{equation}
p= \tfrac{1}{2}qk +r -1, 
\quad
C_1 = h_1^{kq} - h_1^q . 
\end{equation}
Then $F(h)$ will be a rational function iff 
$p$ is an integer, 
and $P(h)$ will be a polynomial of degree $d\leq 4$ iff 
$2r<2r+q<2p+2\leq 4$ are integers. 
Note we can assume $r=0$ or $r=1/2$. 
This determines the following classification:
\begin{subequations}
\begin{align}
2p+2 & = 2,\quad 4,\quad 4,\quad 4,\quad 4,\quad 4
\\
2r +q & = 1,\quad 1,\quad 2,\quad 3,\quad 2,\quad 3
\\
2r & = 0,\quad 0,\quad 0,\quad 0,\quad 1,\quad 1
\end{align}
\end{subequations}
where the corresponding values of $d=2p+2$ and $k=2(p+1-r)/q$ are given by
\begin{align}
d & = 2,\quad 4,\quad 4,\quad 4,\quad 4,\quad 4
\\
k & = 2,\quad 4,\quad 2,\quad \tfrac{4}{3},\quad 3,\quad \tfrac{3}{2}
\end{align}
The only case that has $d\leq 2$ is when $k=2$,
where $q=1$. 
This yields $F(h)=1$ and $P(h)=(h_1-h)(h+h_1-1)$, 
from which the solution integral \eqref{eq:integral-form} is given by 
\begin{equation}
\int_{0}^{h}\frac{F(h)}{\sqrt{P(h)}}\,dh 
= \arctan\bigg( \frac{\sqrt{P(h)}+(2h-1)\sqrt{P(0)}}{2\sqrt{P(h)P(0)}-(h-\frac{1}{2})} \bigg) .
\end{equation}

When $C_1=0$, 
we have 
\begin{gather}
F(h)=  q h^p
\\
P(h) = 1 - h^{2p+2}
\end{gather}
with
\begin{equation}
p= \tfrac{1}{2}q(k-1)-1 . 
\end{equation}
Then $F(h)$ will be a rational function iff 
$p$ is an integer, 
and $P(h)$ will be a polynomial of degree $d\leq 4$ iff 
$0<2p+2\leq 4$ are integers. 
Since $p$ determines $q= 2(p+1)/(k-1)$, 
and $q$ does appear in the powers of the terms in $F(h)$ and $P(h)$, 
we can assume $p=0$ without loss of generality. 
This determines $q= 2/(k-1)$ and $d=2$, 
where $k>1$ is arbitrary. 
Then we have $F(h)=\tfrac{2}{k-1}$ and $P(h) = 1-h^2$. 
Hence, the solution integral \eqref{eq:integral-form} is given by 
\begin{equation}
\int_{0}^{h}\frac{F(h)}{\sqrt{P(h)}}\,dh 
= \frac{2}{(k-1)} \arcsin(h)  . 
\end{equation}

Hence, we obtain the following result,
after taking into account the necessary and sufficient condition \eqref{eq:limiting-g1}
on $g_1$. 

\begin{prop}\label{prop:nodal}
(i) The periodic nodal wave solution integral \eqref{eq:limiting_ode_integral} 
has an explicit evaluation in terms of elementary functions when (and only when) 
$k=2$ with $C_1\neq 0$, 
and $k>1$ arbitrary with $C_1=0$. 
These solutions are given by the explicit expressions
\begin{equation}
\begin{aligned}
k=2:\quad& 
g= \tfrac{1}{2}- \tfrac{1}{2}\cos(\xi-L\lfloor \tfrac{1}{2} + \xi/L\rfloor)+\sqrt{g_1(g_1-1)}\sin(|\xi-L\lfloor \tfrac{1}{2} + \xi/L\rfloor|) , 
\quad
g_1\geq 1, 
\\& L=\pi +2\arctan\bigg(\frac{1}{2\sqrt{g_1(g_1-1)}}\bigg)
\end{aligned}
\label{eq:nodalsol1}
\end{equation}
and 
\begin{equation}
\begin{aligned}
k>1 :\quad& 
g= \sin\big(\tfrac{1}{2}(k-1)|\xi-L\lfloor \tfrac{1}{2} + \xi/L\rfloor|\big)^{2/(k-1)}, 
\\& L=\frac{2\pi}{k-1}
\end{aligned}
\label{eq:nodalsol2}
\end{equation}
where $L$ denotes the wavelength, 
and $\lfloor x \rfloor$ denotes the floor function. \\
(ii) The periodic nodal wave solution integral \eqref{eq:limiting_ode_integral} 
with $C_1\neq 0$ 
has an explicit evaluation in terms of elliptic functions when (and only when) 
$k=\tfrac{4}{3},\tfrac{3}{2},3,4$. 
\end{prop}

\begin{figure}[!h]
\centering
\includegraphics[width=\textwidth]{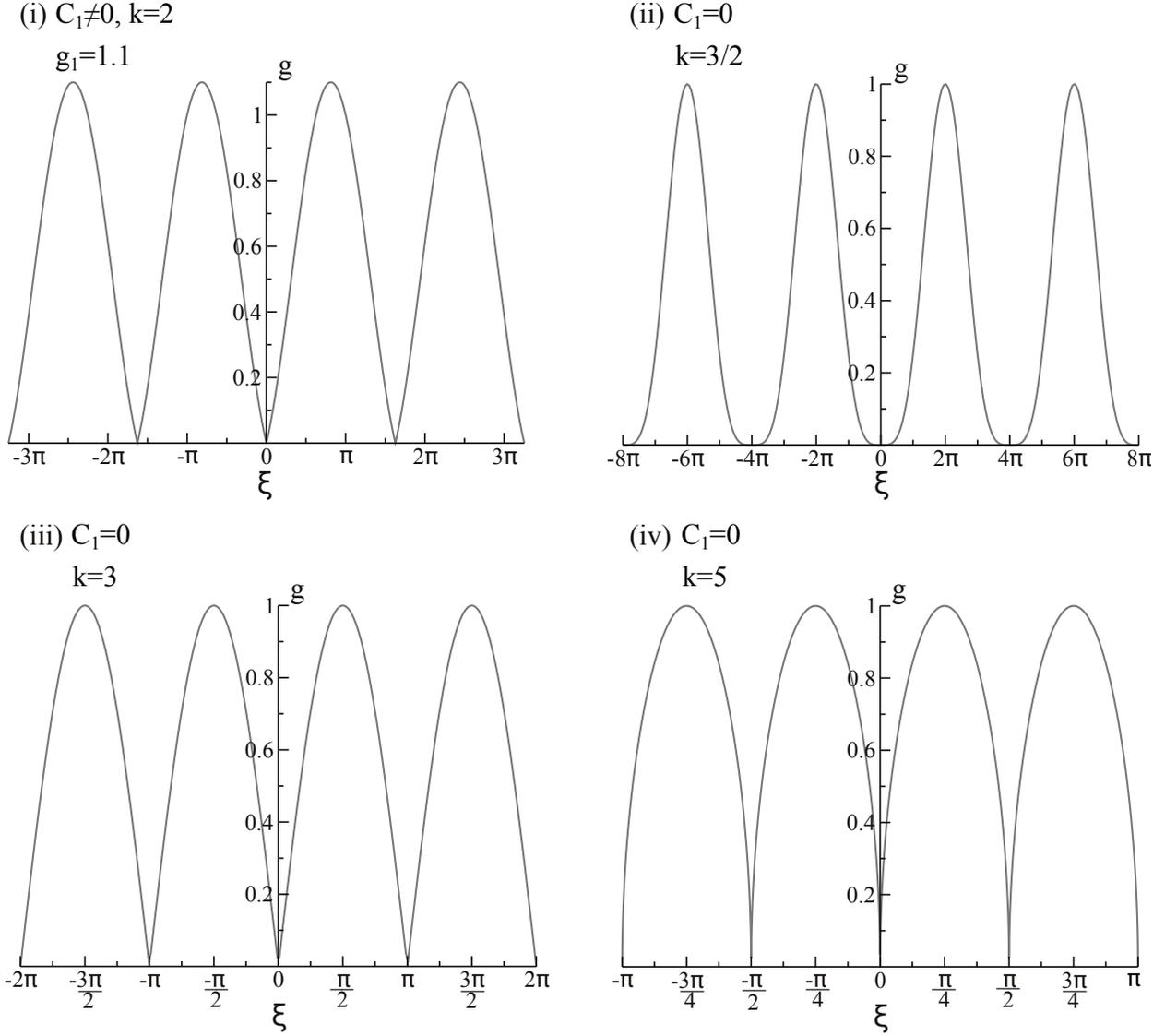}
\caption{Exact periodic nodal wave solutions.}
\label{fig:Nodal-solutions}
\end{figure}

\subsection{Properties of explicit solutions}

For each of the explicit solutions derived in Propositions~\ref{prop:solitary} to~\ref{prop:nodal}, 
we now exhibit their total energy and total momentum 
given by the integrals \eqref{eq:periodic-P}--\eqref{eq:periodic-E}
in the case of periodic waves 
and \eqref{eq:solitary-P}--\eqref{eq:solitary-EP}
in the case of solitary waves. 
These integrals can be evaluated explicitly, 
by the same changes of variable used to evaluate the solution integrals. 

\subsubsection{Solitary waves} 
From Proposition~\ref{prop:solitary}, 
we have two explicit solitary wave solutions \eqref{eq:solitarysol1} and \eqref{eq:solitarysol2}. 
The momentum \eqref{eq:solitary-P}, energy \eqref{eq:solitary-E},
and energy-momentum \eqref{eq:solitary-EP} 
for the first solution \eqref{eq:solitarysol1} are respectively given by 
\begin{align}
&\begin{aligned}
\tilde \P 
%& = 3\sqrt{3}(\nu/c)^4 \nu R \int_{h_0}^{h_1} \frac{h(h+h_0)}{\sqrt{h(1-2h_0-h)}} dh \\&
= (9/2)(\nu/c)^4\nu R \bigg( \big(g_0^2-2g_0+\tfrac{3}{4} \big) \bigg (  \tfrac{1}{2}\pi + \arcsin\bigg(\frac{1-4g_0}{1-2g_0}\bigg)     \bigg) + \tfrac{3}{2}\sqrt{g_0(1-3g_0)} \bigg) ,
\end{aligned}
\label{eq:solitarymom1}
\\
&\begin{aligned}
\tilde\E &
% = (9/2) (\nu/c)^4 \nu^2 R \int_{h_0}^{h_1} \frac{h(2h^2+2hh_0+2h_0-h_0^2)}{\sqrt{h(1-2h_0-h)}} dh \\&
= (9/4)(\nu/c)^4\nu^2 R \bigg( (1-2g_0)\big(g_0^2-\tfrac{3}{2}g_0+\tfrac{5}{4} \big) \bigg (  \tfrac{1}{2}\pi + \arcsin\bigg(\frac{1-4g_0}{1-2g_0}\bigg)  \bigg)
\\&\qquad
+\big(g_0^2-\tfrac{2}{3}g_0+\tfrac{5}{4}\big)\sqrt{g_0(1-3g_0)} \bigg) ,
\end{aligned}
\label{eq:solitaryener1}
\\
&\begin{aligned}
\hat\E &
% = (9/2) (\nu/c)^4 \nu^2 R \int_{h_0}^{h_1} \frac{h(2h^2-h+hh_0+h_0-2h_0^2)}{\sqrt{h(1-2h_0-h)}} dh \\&
= (9/8)(\nu/c)^4\nu^2 R \big(1-\tfrac{3}{2}g_0 \big) \bigg( (1-2g_0)^2 \bigg (  \tfrac{1}{2}\pi + \arcsin\bigg(\frac{1-4g_0}{1-2g_0}\bigg)  \bigg) 
\\&\qquad
+ 2\big(1-\tfrac{4}{3}g_0\big)\sqrt{g_0(1-3g_0)} \bigg) . 
\end{aligned}
\label{eq:solitaryenermom1}
\end{align}

For the second solution \eqref{eq:solitarysol2}, 
the momentum \eqref{eq:solitary-P}, energy \eqref{eq:solitary-E},
and energy-momentum \eqref{eq:solitary-EP} are respectively given by 
\begin{align}
&\begin{aligned}
\tilde\P 
%&= 6\sqrt{2}(\nu/c)^2\nu R \int_{h_0}^{h_1} \frac{h(h + h_0)}{\sqrt{1-h^2-2hh_0-3h_0^2}} dh \\&
= 3\sqrt{2}(\nu/c)^2 \nu R \big(1-2g_0^2\big) \bigg( \tfrac{1}{2}\pi - \arcsin\bigg(\frac{2g_0}{\sqrt{1-2g_0^2}} \bigg) \bigg),
\end{aligned}
\label{eq:solitarymom2}
\\
&\begin{aligned}
\tilde\E 
%&= 6\sqrt{2}(\nu/c)^2\nu^2 R \int_{h_0}^{h_1} \frac{h \big( (h+h_0)(h^2+h_0^2) +h_0(1-2h_0^2) \big)}{\sqrt{1-h^2 -2hh_0-3h_0^2}} dh \\&
= (9/4)\sqrt{2}(\nu/c)^2 \nu^2 R \big(1-2g_0^2\big) \bigg( \big(1+\tfrac{2}{3}g_0^2\big) \bigg(\tfrac{1}{2}\pi - \arcsin\bigg(\frac{2g_0}{\sqrt{1-2g_0^2}} \bigg) \bigg)
-\tfrac{2}{3}g_0\sqrt{1-6g_0^2} \bigg),
\end{aligned}
\label{eq:solitaryener2}
\\
&\begin{aligned}
\hat\E 
%&= 3\sqrt{2}(\nu/c)^2\nu^2 R \int_{h_0}^{h_1} \frac{h (2h^2+2hh_0+h_0^2-1) +h_0(1-3h_0^2)}{\sqrt{1-h^2 -2hh_0-3h_0^2}} dh \\&
= (3/4)\sqrt{2}(\nu/c)^2 \nu^2 R \big(1-2g_0^2\big) \bigg( \tfrac{1}{2}\pi - \arcsin\bigg(\frac{2g_0}{\sqrt{1-2g_0^2}} \bigg)  -2g_0\sqrt{1-6g_0^2} \bigg) . 
\end{aligned}
\label{eq:solitaryenermom2}
\end{align}

The preceding expressions are plotted as functions of $g_0$ in Fig.~\ref{fig:energy-momentum}. 

\subsubsection{Cusped solitary waves} 
From Proposition~\ref{prop:cuspedsolitary}, 
we have two explicit cusped solitary wave solutions \eqref{eq:cuspedsolsol1} and \eqref{eq:cuspedsolsol2}, 
The momentum \eqref{eq:solitary-P}, energy \eqref{eq:solitary-E},
and energy-momentum \eqref{eq:solitary-EP} 
for the first solution \eqref{eq:cuspedsolsol1} are respectively given by 
\begin{align}
&\begin{aligned}
-\tilde\P 
%&= 9(\nu/c)^4 \nu R \int_{0}^{h_0} \frac{h(h+h_0)}{\sqrt{h(1-2h_0-h)}} dh \\&
= (9/2)(\nu/c)^4\nu R \bigg( \big(g_0^2-2g_0+\tfrac{3}{4} \big) \bigg (  \tfrac{1}{2}\pi - \arcsin\bigg(\frac{1-4g_0}{1-2g_0}\bigg)     \bigg) - \tfrac{3}{2}\sqrt{g_0(1-3g_0)} \bigg) ,
\end{aligned}
\label{eq:cuspedsolmom1}
\\
&\begin{aligned}
-\tilde\E &
%= (9/2) (\nu/c)^4 \nu^2 R \int_{0}^{h_0} \frac{h(2h^2+2hh_0+2h_0-h_0^2)}{\sqrt{h(1-2h_0-h)}} dh \\&
= (9/4)(\nu/c)^4\nu^2 R \bigg( (1-2g_0)\big(g_0^2-\tfrac{3}{2}g_0+\tfrac{5}{4} \big) \bigg (  \tfrac{1}{2}\pi - \arcsin\bigg(\frac{1-4g_0}{1-2g_0}\bigg)  \bigg)
\\&\qquad
-2 \big(g_0^2-\tfrac{2}{3}g_0+\tfrac{5}{4}\big)\sqrt{g_0(1-3g_0)} \bigg) ,
\end{aligned}
\label{eq:cuspedsolener1}
\\
&\begin{aligned}
-\hat\E & 
% = (9/2) (\nu/c)^4 \nu^2 R \int_{0}^{h_0} \frac{h(2h^2-h+hh_0+h_0-2h_0^2)}{\sqrt{h(1-2h_0-h)}} dh\\
= (9/8)(\nu/c)^4\nu^2 R \big(1-\tfrac{3}{2}g_0 \big) \bigg( (1-2g_0)^2 \bigg (  \tfrac{1}{2}\pi - \arcsin\bigg(\frac{1-4g_0}{1-2g_0}\bigg)  \bigg) 
\\&\qquad
- 2\big(1-\tfrac{4}{3}g_0\big)\sqrt{g_0(1-3g_0)} \bigg) . 
\end{aligned}
\label{eq:cuspedsolenermom1}
\end{align}
Note that, compared to the case of solitary waves,
the overall negative sign in each of these expressions 
is due to the domain of integration being $0\leq h \leq h_{\max}$, 
rather than $h_{\min}\leq h \leq h_{\max}$,  

Similarly, 
the momentum \eqref{eq:solitary-P}, energy \eqref{eq:solitary-E},
and energy-momentum \eqref{eq:solitary-EP} 
for the second solution \eqref{eq:cuspedsolsol2} are respectively given by 
\begin{align}
&\begin{aligned}
-\tilde\P & 
%= 6\sqrt{2}(\nu/c)^2\nu R \int_{0}^{h_0} \frac{h(h + h_0)}{\sqrt{1-h^2-2hh_0-3h_0^2}} dh \\&
= 3\sqrt{2}(\nu/c)^2 \nu R \bigg( \big(1-2g_0^2\big) \bigg( \arcsin\bigg(\frac{2g_0}{\sqrt{1-2g_0^2}} \bigg) - \arcsin\bigg(\frac{g_0}{\sqrt{1-2g_0^2}} \bigg) \bigg) 
\\&\qquad
-g_0\sqrt{1-3g_0^2} \bigg),
\end{aligned}
\label{eq:cuspedsolmom2}
\\
&\begin{aligned}
-\tilde\E & 
%= 6\sqrt{2}(\nu/c)^2\nu^2 R \int_{0}^{h_0} \frac{h \big( (h+h_0)(h^2+h_0^2) +h_0(1-2h_0^2) \big)}{\sqrt{1-h^2 -2hh_0-3h_0^2}} dh \\
= 3\sqrt{2}(\nu/c)^2 \nu^2 R \bigg( \big(g_0^4 +g_0^2-\tfrac{3}{4}\big) \bigg( \arcsin\bigg(\frac{g_0}{\sqrt{1-2g_0^2}} \bigg) 
-\arcsin\bigg(\frac{2g_0}{\sqrt{1-2g_0^2}} \bigg) \bigg)
\\&\qquad
 +g_0 \bigg( \tfrac{5}{4}\big(1-\tfrac{4}{5}g_0^2\big)\sqrt{1-3g_0^2}-\tfrac{1}{2}\big(1-2g_0^2\big)\sqrt{1-6g_0^2} \bigg) \bigg),
\end{aligned}
\label{eq:cuspedsolener2}
\\
&\begin{aligned}
-\hat\E & 
%= 3\sqrt{2}(\nu/c)^2\nu^2 R \int_{0}^{h_0} \frac{h (2h^2+2hh_0+h_0^2-1) +h_0(1-3h_0^2)}{\sqrt{1-h^2 -2hh_0-3h_0^2}} dh \\
= (3/4)\sqrt{2}(\nu/c)^2 \nu^2 R \big(1-2g_0^2\big) \bigg(  \arcsin\bigg(\frac{2g_0}{\sqrt{1-2g_0^2}} \bigg) - \arcsin\bigg(\frac{g_0}{\sqrt{1-2g_0^2}} \bigg)  
\\&\qquad
+g_0 \bigg( 2\sqrt{1-6g_0^2} -3\sqrt{1-3g_0^2} \bigg) \bigg) .
\end{aligned}
\label{eq:cuspedsolenermom2}
\end{align}

\subsubsection{Periodic waves} 
From Proposition~\ref{prop:periodic}, 
we have two explicit periodic wave solutions \eqref{eq:periodicsol1} and \eqref{eq:periodicsol2}.
The momentum \eqref{eq:periodic-P} and energy~\eqref{eq:periodic-E}
are given by 
\begin{align}
\begin{aligned}
\P
%&= 9(\nu/c)^4\nu R \int_{h_0}^{h_1} \frac{h^2}{\sqrt{(h_1-h)(h-1+h_1)}}dh
=(9/2)(\nu/c)^4\pi\nu R \big(g_1^2-g_1+\tfrac{3}{4}\big),
\end{aligned}
\label{eq:periodicmom1}
\\
\begin{aligned}
\E
%&= (9/2)(\nu/c)^4\nu^2 R \int_{h_0}^{h_1} \frac{h(2h^2+h_1-h_1^2)}{\sqrt{(h_1-h)(h-1+h_1)}}dh
=(9/2)(\nu/c)^4\pi\nu^2 R \big(g_1^2-g_1+\tfrac{5}{8}\big)
\end{aligned}
\label{eq:periodicener1}
\end{align}
for the first solution,
and likewise 
\begin{align}
&\begin{aligned}
\P
%&= 3\sqrt{2}(\nu/c)^2\nu R \int_{h_0}^{h_1} \frac{h}{\sqrt{(h_1-h)(h-1+h_1)}}dh 
= \big(3/\sqrt{2}\big) (\nu/c)^2 \pi \nu R , 
\end{aligned}
\label{eq:periodicmom2}
\\
&\begin{aligned}
\E 
%&= \big(3/\sqrt{2}\big)(\nu/c)^2\nu^2 R \int_{h_0}^{h_1} \frac{2h^2+h_1-h_1^2}{\sqrt{(h_1-h)(h-1+h_1)}}dh
=\big(9\sqrt{2}/8\big)(\nu/c)^2\pi\nu^2 R 
\end{aligned}
\label{eq:periodicener2}
\end{align}
for the second solution.

\begin{figure}[!h]
\centering
\includegraphics[width=\textwidth]{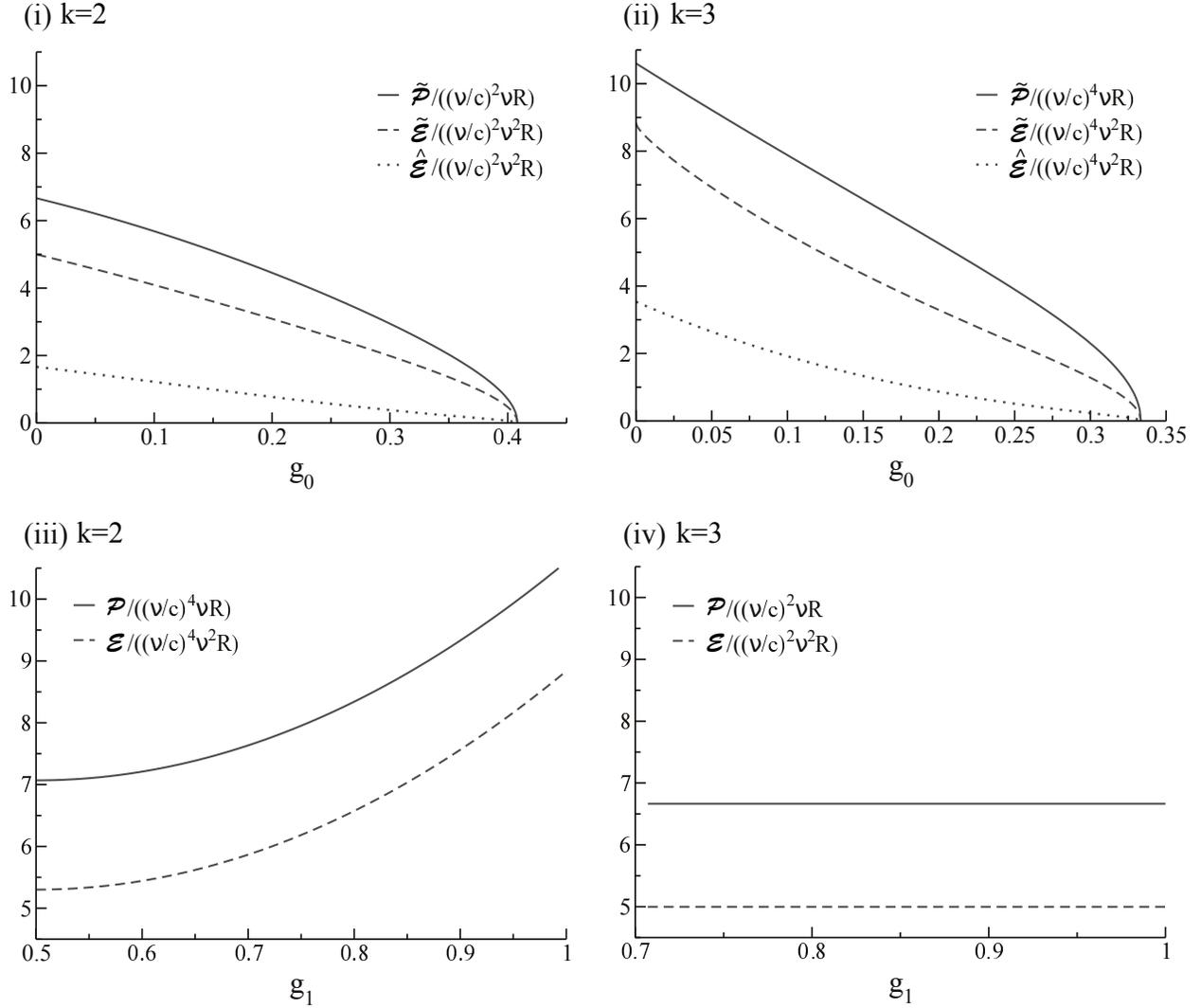}
\caption{(i), (ii): Energy and momentum for solitary waves. (iii), (iv): Energy and momentum for periodic waves.}
\label{fig:energy-momentum}
\end{figure}

The preceding expressions as functions of $g_1$ are illustrated in Fig.~\ref{fig:energy-momentum}. 

\subsubsection{Cusped periodic waves} 
From Proposition~\ref{prop:cuspedperiodic}, 
we have one cusped periodic wave solution \eqref{eq:cuspedpersol1}. 
Its momentum \eqref{eq:periodic-P} and energy \eqref{eq:periodic-E}
are given by 
\begin{align}
&\begin{aligned}
\P
%&= 3\sqrt{2}(\nu/c)^2\nu R \int_{0}^{h_0} \frac{h}{\sqrt{(h_0-h)(h-1+h_0)}}dh \\&
= \big(3/\sqrt{2}\big) (\nu/c)^2 \nu R \bigg( \tfrac{1}{2}\pi + \arcsin\bigg(\frac{1}{2g_0^2-1}\bigg) + 2\sqrt{g_0^2(g_0^2-1)}\bigg) , 
\end{aligned}
\label{eq:cuspedpermom1}
\\
&\begin{aligned}
\E
%&= \big(3/\sqrt{2}\big)(\nu/c)^2\nu^2 R \int_{0}^{h_0} \frac{2h^2+h_0-h_0^2}{\sqrt{(h_0-h)(h-1+h_0)}}dh\\&
=\big(9\sqrt{2}/8\big)(\nu/c)^2\nu^2 R \bigg( \tfrac{1}{2}\pi + \arcsin\bigg(\frac{1}{2g_0^2-1}\bigg) + 2\sqrt{g_0^2(g_0^2-1)}\bigg).
\end{aligned}
\label{eq:cuspedperener1}
\end{align}

\subsubsection{Nodal periodic waves} 
From Proposition~\ref{prop:nodal}, 
we have two explicit nodal periodic wave solutions
\eqref{eq:nodalsol1} and \eqref{eq:nodalsol2}. 
The momentum \eqref{eq:periodic-P} and energy \eqref{eq:periodic-E}
for the first solution are given by 
\begin{align}
&\begin{aligned}
\P
%&= 9 (\nu/c)^4\nu R \int_0^{h_1}\frac{h^2}{\sqrt{(h_1-h)(h-1+h_1)}}dh \\&
= (9/2) (\nu/c)^4\nu R \bigg ( \big(g_1^2-g_1+\tfrac{3}{4}\big) \bigg( \tfrac{1}{2}\pi + \arcsin \bigg( \frac{1}{2g_1-1}\bigg) \bigg) + \tfrac{3}{2}\sqrt{g_1(g_1-1)} \bigg), 
\end{aligned}
\label{eq:nodalmom1}
\\
&\begin{aligned}
\E
%&= (9/2) (\nu/c)^4\nu^2 R \int_0^{h_1}\frac{h(2h^2+h_1-h_1^2)}{\sqrt{(h_1-h)(h-1+h_1)}}dh \\&
& = (9/2) (\nu/c)^4\nu^2 R \bigg ( \big(g_1^2-g_1+\tfrac{5}{8}\big) \bigg( \tfrac{1}{2}\pi + \arcsin \bigg( \frac{1}{2g_1-1}\bigg) \bigg) 
\\&\qquad
+ \tfrac{1}{3}(g_1^2-g_1+\tfrac{15}{4})\sqrt{g_1(g_1-1)} \bigg) . 
\end{aligned}
\label{eq:nodalener1}
\end{align}
Similarly, 
the momentum \eqref{eq:periodic-P} and energy \eqref{eq:periodic-E}
for the second solution are given by 
\begin{align}
&\begin{aligned}
\P
%&= (2/\sqrt{3}) (\nu/c)^{4/(k-1)}\nu R (k-1)^{-1} \sqrt{k(k+1)/2}^{(k+3)/(k-1)} \int_0^1 \frac{h^{4/(k-1)}}{\sqrt{1-h^2}}dh \\&
=\sqrt{\pi/3} (\nu/c)^{4/(k-1)}\nu R (k-1)^{-1} \sqrt{k(k+1)/2}^{(k+3)/(k-1)} \Gamma\big(\tfrac{k+3}{2(k-1)} \big)/
\Gamma\big(\tfrac{k+1}{k-1} \big),
\end{aligned}
\label{eq:nodalmom2}
\\
&\begin{aligned}
\E
%&= (2/\sqrt{3}) (\nu/c)^{4/(k-1)}\nu^2 R (k-1)^{-1} \sqrt{k(k+1)/2}^{(k+3)/(k-1)} \int_0^1 \frac{h^{2(k+1)/(k-1)}}{\sqrt{1-h^2}}dh \\&
=\sqrt{\pi/3} (\nu/c)^{4/(k-1)}\nu^2 R (k-1)^{-1} \sqrt{k(k+1)/2}^{(k+3)/(k-1)} \Gamma\big(\tfrac{3k+1}{2(k-1)} \big)/
\Gamma\big(\tfrac{2k}{k-1} \big) ,
\end{aligned}
\label{eq:nodalener2}
\end{align}
where $\Gamma(x)$ denotes the gamma function. 

\section{Concluding Remarks}\label{sec:remarks}

For the highly nonlinear, fourth-order wave equation \eqref{eq:tilu-pde}, 
we have obtained all solitary wave solutions and all nonlinear periodic solutions. 
We are able to parameterize these solutions explicitly in terms of 
the asymptotic value of the wave amplitude in the case of solitary waves
and the peak of the wave amplitude in the case of nonlinear periodic waves. 
All cases in which the solution expressions can be stated in an explicit analytic form 
using elementary functions are worked out. 
Altogether, this yields 
two explicit solitary waves, two explicit periodic waves, 
three explicit cusped waves, 
and two explicit nodal waves. 
We also give explicit expressions for the total energy and total momentum 
for each of these solutions. 

Our method for deriving the solutions has some novel aspects 
compared to the standard travelling wave ansatz. 
In particular, firstly, 
we use the conservation laws admitted by 
the highly nonlinear, fourth-order wave equation \eqref{eq:tilu-pde}
to reduce it directly to a separable first-order ODE 
for travelling waves. 
Secondly, 
we classify all possible types of travelling wave solutions by
a modified energy analysis argument. 
Thirdly, 
we use a systematic method to determine all cases of the nonlinearity exponent $k>1$ 
for which each of the different travelling wave solutions
can be evaluated in an explicit analytical form in terms of elementary functions. 
Fourthly, 
we show how to use the ODE along with its first integrals 
to obtain the total energy and total momentum of the travelling wave solutions 
without the need to evaluate complicated integrals involving their explicit form. 
This method can be applied more broadly to any quasilinear wave equation.

\section*{Acknowledgments}
We are grateful to Surajit Sen and Thad Harroun for their enthusiasm and interest in this work. 
S.C.A.\ is supported by an NSERC research grant.
M.P.\ is supported by a Vanier Canada Graduate Scholarship from NSERC.

\end{document}